\documentclass[prx,superscriptaddress,amsmath,amssymb,twocolumn,nonnormative]{revtex4-2}
\usepackage{graphicx,bm,amsmath}
\usepackage[usenames,dvipsnames]{xcolor}
\usepackage[colorlinks,bookmarks=false,citecolor=blue,linkcolor=red,urlcolor=blue,
]{hyperref}
\usepackage[percent]{overpic}
\usepackage[bbgreekl]{mathbbol}
\usepackage{xcolor}
\usepackage[normalem]{ulem}
\usepackage{algorithm}
\usepackage{braket}
\usepackage{algpseudocode}
\usepackage{comment}
\usepackage{amsthm}
\usepackage{enumerate}
\usepackage{subfigure}
\usepackage{lipsum}
\usepackage{orcidlink}
\usepackage{subfigure}
\usepackage{booktabs}  
\def\doi{http://dx.doi.org/}
\newcommand{\be}{\begin{equation}}
\newcommand{\ee}{\end{equation}}
\newcommand{\bec}{\begin{equation*}}
\newcommand{\eec}{\end{equation*}}
\newcommand{\bea}{\begin{eqnarray}}
\newcommand{\eea}{\end{eqnarray}}

\newcommand{\Tr}{\text{Tr}}   

\newcommand{\dket}[1]{\left|#1\right\rangle\!\rangle}

\newcommand{\titleinfo}{Mixed-State Phase Transitions in Measurement-Dressed Imaginary-Time Evolution}
\begin{document}
\title{\titleinfo}
    
\author{Yi-Ming Ding}
\affiliation{Department of Physics, School of Science and Research Center for Industries of the Future, Westlake University, Hangzhou 310030,  China}
\affiliation{Institute of Natural Sciences, Westlake Institute for Advanced Study, Hangzhou 310024, China}
\affiliation{State Key Laboratory of Surface Physics and Department of Physics, Fudan University, Shanghai 200438, China}

\author{Zenan Liu}
\email{liuzenan@westlake.edu.cn}
\affiliation{Department of Physics, School of Science and Research Center for Industries of the Future, Westlake University, Hangzhou 310030,  China}
\affiliation{Institute of Natural Sciences, Westlake Institute for Advanced Study, Hangzhou 310024, China}

\author{Xu Tian}
\affiliation{Department of Physics, School of Science and Research Center for Industries of the Future, Westlake University, Hangzhou 310030,  China}
\affiliation{Institute of Natural Sciences, Westlake Institute for Advanced Study, Hangzhou 310024, China}

\author{Zhe Wang}
\affiliation{Department of Physics, School of Science and Research Center for Industries of the Future, Westlake University, Hangzhou 310030,  China}
\affiliation{Institute of Natural Sciences, Westlake Institute for Advanced Study, Hangzhou 310024, China}

\author{Yanzhang Zhu}
\affiliation{State Key Laboratory of Surface Physics and Department of Physics, Fudan University, Shanghai 200438, China}
\affiliation{Department of Physics, School of Science and Research Center for Industries of the Future, Westlake University, Hangzhou 310030,  China}
\affiliation{Institute of Natural Sciences, Westlake Institute for Advanced Study, Hangzhou 310024, China}

\author{Zheng Yan}
\email{zhengyan@westlake.edu.cn}
\affiliation{Department of Physics, School of Science and Research Center for Industries of the Future, Westlake University, Hangzhou 310030,  China}
\affiliation{Institute of Natural Sciences, Westlake Institute for Advanced Study, Hangzhou 310024, China}

\begin{abstract}
Motivated by the ubiquity of decoherence in quantum hardware and the growing role of imaginary-time evolution (ITE) in quantum algorithms, we investigate how many-body correlations generated by imaginary-time filtering are modified by local decoherence. 
We introduce \emph{measurement-dressed imaginary-time evolution} (MDITE), which alternates ITE with projective-measurement channels, producing a competition between low-energy filtering and local dephasing. 
By developing a new efficient quantum Monte Carlo method, we uncover MDITE mixed-state transitions with spontaneous-symmetry-breaking signatures in the driving of 1D transverse-field Ising and 2D columnar dimerized Heisenberg Hamiltonians in the resulting density matrices.
In the continuous limit, the Choi--Jamiołkowski mapping yields a tractable equilibrium description with conformal criticality that qualitatively captures the phase transitions. At finite protocol parameters, however, the four-point correlator violates the conformal cross-ratio form and the critical exponents deviate from their continuous-limit values, signaling the loss of conformal symmetry and richer nonequilibrium criticality.
Our results establish MDITE as a controlled setting for exploring mixed-state phases and critical phenomena driven by the interplay between imaginary-time filtering and decoherence.
\end{abstract}

\maketitle

\section{Introduction}
Decoherence is one of the central mechanisms by which quantum systems lose quantum coherence through their interaction with environmental degrees of freedom
~\cite{Zeh1970decoherence,Zurek198pointer,Zurek1982superselection,Zurek2001EnvInduced,Zurek2003review,Schlosshauer2005review,cai2013algebraic,cai2014identifying,yan2018interacting,Schlosshauer2019review}.
It converts coherent superpositions into statistical mixtures in a stable pointer
basis, suppresses quantum interference, and provides a microscopic route toward
effectively classical behavior~\cite{Zurek198pointer,Zurek1982superselection}.
In quantum many-body systems, decoherence is not only a practical obstacle for
quantum technologies, where it destroys the coherence and
entanglement required for quantum advantage
~\cite{NielsenChuang2010,Bennett1996purification,Bravyi2005universal,Gottesman2009errorcorrection,Terhal2015errorcorrection},
but also a fundamental mechanism that can reorganize correlations, phases, and
critical behavior in mixed states
~\cite{cai2013algebraic,cai2014identifying,yan2018interacting,Merkli2007Decoherence}.
This motivates the study of quantum many-body physics directly at the level of density
matrices and quantum channels, rather than only through isolated pure-state
dynamics.

Meanwhile, imaginary-time evolution (ITE) offers another fundamental
non-unitary mechanism in quantum many-body physics: replacing the real-time
evolution operator $e^{-itH}$ by $e^{-\tau H}$ filters a state toward the
low-energy sector of the Hamiltonian $H$.
Its filtering property underlies many computational approaches, including
quantum Monte Carlo (QMC) and tensor-network algorithms
~\cite{Vidal2007Classical,jiang20082dtnprojector,Jordan2008tn-imag,Haegeman2016tn-imag,sandvik2003stochastic,Prokofev1998wordline,Sandvik1992sse,Jiang2008Accurate,Grandi2011Universal,Zi2019Sign,Assaad2008worldline,Liu2013Quasi}.
Beyond its computational utility, imaginary-time dynamics also provides a useful framework for probing quantum critical behavior both in and out of equilibrium
~\cite{Grandi2011Universal,Yin2014Universal,Liu2013Quasi,Shu2022Nonequilibrium,Yu2026Nonequilibrium,Yin2026Preempting,wang2026noncommutative}.
More recently, advances in quantum algorithms and programmable quantum platforms have elevated ITE from a formal theoretical construction to an experimentally relevant operation~\cite{Motta2020qite,McArdle2019vqite,Yuan2019theoryofvariational,Benedetti2021qite,Kondappan2023qite,Mao2023qite,Ding2024qite,Nishi2021qite-nisq,Cao2022qite,ZhangSX2024qite,Semeghini2021Probing,Lin2021QITE}. Because realistic quantum devices are inevitably subject to noise and decoherence, a natural question is how these environmental effects modify the many-body correlations and phases generated by imaginary-time filtering.

In this work, we address this question in a minimal channel-based setting, where imaginary-time filtering competes with decoherence generated by local projective measurements. ITE suppresses high-energy components and
builds low-energy correlations associated with the Hamiltonian, while the
projective-measurement channel dephases the density matrix in the local
measurement basis. This setup is motivated in part by the broader lesson of
measurement-induced critical phenomena: measurements, when combined with
many-body dynamics, can qualitatively reorganize correlations and information
structure ~\cite{liyaodong2018clifford,liyaodong2019clifford,Skinner2019,fisher2023randomcircuit}.
Here, however, ``measurement'' refers to the outcome-averaged quantum channel. Accordingly, the central object of our study is the resulting mixed-state density matrix, rather than an ensemble of postselected pure-state trajectories.

We investigate this problem through a protocol that we call
\emph{measurement-dressed imaginary-time evolution} (MDITE). 
starting from the maximally mixed state, the protocol alternates between ITE generated by a local Hamiltonian and local outcome-averaged projective measurement channels. The maximally mixed state serves as an unbiased
infinite-temperature reference, so any ordering or criticality in the resulting mixed state is generated by the protocol itself rather than inherited from the initial state. 
In the absence of measurements, the protocol reduces to imaginary-time filtering toward the low-energy sector of the Hamiltonian; in the presence of measurements, local dephasing competes with this filtering process. 
As we show below, this competition gives rise to sharp mixed-state phase transitions controlled by the imaginary evolution time, the measurement strength, and the parameters of the Hamiltonian.

Because the nonequilibrium mixed states generated by MDITE are difficult to treat analytically, we develop a diagrammatic representation of the protocol that enables its efficient incorporation into QMC simulations. 
The main text focuses on the physical phase structure and its interpretation, the technical details of the diagrammatic construction and QMC algorithm are presented in Appendices~\ref{appx:pathint} and~\ref{appx:qmc2}.
Concretely, we demonstrate mixed-state phase transitions of MDITE in two representative many-body local Hamiltonians: the 1D transverse-field Ising model and the 2D columnar dimerized Heisenberg model. In both cases, projective measurements are performed in the computational basis and averaged over all outcomes. 
We reveal that the MDITE steady states exhibit sharp transitions accompanied by numerical signatures of $\mathbb{Z}_2$ weak-to-trivial spontaneous symmetry breaking (SSB). Accordingly, these transitions can be diagnosed using conventional linear order parameters, such as the spin magnetization, rather than the nonlinear information-theoretic diagnostics commonly employed in trajectory-resolved measurement-induced transitions. Their phase structure is therefore directly accessible at the level of the mixed-state density matrix and is well suited to experimental detection.

An important feature of MDITE, as discussed in Sec.~\ref{sec:protocol}, is that its control parameters including the imaginary-time step, Hamiltonian couplings, and measurement rates, may take arbitrary finite values. 
This is consistent with the fact that general realistic experimental noise processes may not always admit a continuous-time description.

To elucidate the underlying mechanism of phase transitions, we first analyze the continuous-parameter limit, where we map the MDITE mixed state onto a doubled Hilbert space through the Choi--Jamiołkowski isomorphism~\cite{Choi1975,Jamiolkowski1972}. This mapping yields a theoretically tractable equilibrium description of the steady state under an effective local Hamiltonian. This reveals how the interplay among measurement, ITE, and Hamiltonian parameters drives transitions between distinct mixed-state phases. 
The transitions in this limit belong to the conventional Ising universality classes and exhibit conformal symmetry. Although derived in a specific limit, this description qualitatively accounts for the phase structure observed numerically at finite protocol parameters, including the emergence of SSB mixed states. It does not, however, quantitatively capture the corresponding finite-parameter universality classes.

Meanwhile, our simulations at finite protocol parameters reveal well-defined transitions with critical exponents that differ substantially from their continuous-limit values. Although the fitted dynamical exponent $z\approx 1$ suggests an emergent equivalence between space and time, the four-point correlation functions cannot be consistently described by the conformal cross-ratio form. This indicates that, away from the continuous-parameter limit, the critical behavior is not governed by a conformal field theory (CFT) and may instead reflect a richer nonequilibrium critical structure.

We remark that, although this work focuses on local projective-measurement channels, they should be viewed as a simple and controlled starting point for studying the interplay between ITE and decoherence. Other choices of Hamiltonians and decoherence channels may therefore give rise to a richer landscape of mixed-state phases and critical phenomena relevant to experimental quantum simulators or quantum computers.

This paper is organized as follows. In Sec.~\ref{sec:protocol}, we formally introduce the MDITE protocol and discuss its key features. In
Secs.~\ref{sec:result:tfim} and \ref{sec:result:cdhm}, we report the numerical results of the 1D transverse-field Ising model and the 2D columnar
dimerized Heisenberg model, respectively, and present the corresponding
mixed-state phases and critical behavior. In Sec.~\ref{sec:theoretical}, we
analyze the continuous-parameter limit in which the dynamics maps onto an equilibrium problem of an effective local Hamiltonian, providing qualitative insight into the mechanism of the observed transitions. 

To keep the main text focused on the physical results and their interpretation, we present several technical developments in the appendices. Appendix~\ref{appx:pathint} introduces a diagrammatic representation of the MDITE process that clarifies the channel structure and enables efficient numerical simulations. 
The corresponding QMC algorithm is presented in Appendix~\ref{appx:qmc2}. An interpretation of the mixed-state transitions in terms of cluster formation during the QMC updates is further discussed in Appendix~\ref{appx:qmc3}.
Finally, in Sec.~\ref{sec:discussion}, we summarize our findings and outline directions for future research.

\section{Protocol}\label{sec:protocol}
In this section, we formally define the \emph{measurement-dressed imaginary-time evolution (MDITE)} protocol. While our discussion focuses on qubit systems, it can be readily generalized to systems with arbitrary local dimensions.

Given a local Hamiltonian $H$ describing a quantum many-body system of $N$ qubits, its associated MDITE protocol is defined as follows:
\begin{enumerate}
    \item \small{\textbf{Input state}}: The protocol starts with the maximally mixed state $\rho_0=I_{2^N}/2^N$, where $I_{2^N}$ is the identity operator of dimension $2^N$.
    \item \small{\textbf{Evolution}}: 
    For each discrete time $k = 1, 2, \dots$, the state evolves through two consecutive operations: 
    \begin{itemize}
        \item A probabilistic projective measurement channel $\mathcal{E}_p$ in the computational basis is applied, where the measurement rate $p$ controls the probability that the qubits are measured;
        \item An ITE with respect to $H$. 
    \end{itemize}
    The resulting state is therefore
    \begin{equation}
        \rho_k = \frac{1}{\mathcal{N}_k}e^{-\frac{\tau}{2}H} \mathcal{E}_p[\rho_{k-1}] e^{-\frac{\tau}{2}H},
    \end{equation}
    where $\mathcal{N}_k = \Tr \!\left(e^{-\frac{\tau}{2} H} \mathcal{E}_p[\rho_{k-1}] e^{-\frac{\tau}{2} H}\right)$  
    is the normalization factor ensuring $\Tr(\rho_k) = 1$.
    \item \small{\textbf{Output state}}: 
    At time $n_d$, the final mixed state $\rho_{n_d}$ is obtained as the output. 
\end{enumerate}

Specifically, we consider that each qubit is independently measured with probability $p \in [0,1]$, and left unchanged with probability $(1-p)$.
Mathematically,
\begin{equation}
    \mathcal{E}_p = \prod_{i=1}^N \mathcal{E}_{p,i},
\end{equation}
with 
\begin{equation}\label{eq:local_channle}
    \begin{aligned}
        \mathcal{E}_{p,i}[\rho] \equiv & (1-p)\rho  \\ & + p \bigg(
    \ket{0_i}\langle 0_i|\rho |0_i\rangle \bra{0_i}    + 
     \ket{1_i}\langle 1_i|\rho |1_i\rangle \bra{1_i}    
    \bigg).
    \end{aligned}
\end{equation}
Here $\ket{0}\equiv\ket{\uparrow}$ and $\ket{1}\equiv\ket{\downarrow}$ denote the eigenstates of the local Pauli operator $Z_i$ acting on the $i$th qubit.

We are interested in the regime where, after a sufficiently large number of time steps $n_d$, the system approaches a steady state, i.e., $\rho_{n_d} \approx \rho_{n_d+1} \approx \rho_{\infty}$. 
However, we stress that the existence of $\rho_{\infty}$ is not guaranteed for general decoherence channels. Nevertheless, for the examples considered in this work, such a steady state indeed exists, as we will show in the following sections.

The dynamical process of MDITE involves a competition between two effects: the ITE operator $e^{-\frac{\tau}{2}H}$, which drives the system toward the ground state of $H$, and the measurement channel $\mathcal{E}_p$, which introduces decoherence and tends to produce a mixed state. 
As the computational basis is generally not the eigenbasis of $H$, ITE induces quantum fluctuations and reduces classical uncertainty. 
This suggests potential mixed-state phase transitions with respect to the steady state $\rho_{\infty}$ when tuning the parameters in the protocol.

It is important to note that the parameter $\tau$ of each ITE in each discrete MDITE time and the measurement rate $p$ serve as primary parameters controlling the behaviors of the steady state $\rho_{\infty}$, as we will discuss in Sec.~\ref{sec:theoretical}.
Moreover, Hamiltonian parameters (e.g., coupling strength) also play a crucial role, resulting in a high-dimensional parameter space.
This enriches the landscape of behaviors of the output state, rendering the setup of MDITE compelling for investigating and discovering fruitful novel many-body physics in open quantum systems. 

\section{Numerical results for a 1D Model with discrete symmetry}\label{sec:result:tfim}
In the following, we discuss two explicit examples with
mixed-state phase transitions in their MDITEs. 
For each model, we present numerical results alongside a detailed analysis, including estimates of the critical exponents and fits of the four-point correlation functions to the conformal cross-ratio form.

As noted above, MDITE mixed states are analytically intractable in generic interacting systems, making numerical simulations indispensable. 
However, existing numerical methods are not directly applicable to ITE with projective measurement channels, especially in higher
spatial dimensions. To overcome this difficulty, we develop a diagrammatic
representation of the MDITE process, which further leads to the QMC algorithm
used to obtain the results reported below. 
Since the algorithmic construction is primarily technical, we defer its details to Appendices~\ref{appx:pathint} and~\ref{appx:qmc2}. For completeness, Appendix~\ref{appx:qmc1} also provides an overview of QMC.

\subsection{Model and observables}
As the first example, we consider the 1D transverse-field Ising model (TFIM)~\cite{pfeuty1970one} as the local Hamiltonian in the MDITE protocol, which is given by 
\begin{equation}\label{eq:h_tfim}
    H=-\sum_{i}Z_iZ_{i+1}-h\sum_i X_i ,
\end{equation}
where $Z_i$ and $X_i$ are the Pauli operators acting on site $i$, and $h$ is the transverse field strength. 
The computational basis is chosen to be the $Z$-basis. 
The length of the chain is denoted by $L\equiv N$, and the periodic boundary condition, i.e., $N+1 \equiv 1$, is considered. 

This model has a global $\mathbb{Z}_2$ symmetry and a ground-state quantum phase transition at the self-dual point $h_{c,\mathrm{GS}}=1$, as determined by Kramers--Wannier duality. The transition belongs to the 2D classical Ising universality class and is described by the $(1+1)$D Ising CFT.
For $h < h_{c,\mathrm{GS}}$, the system is in a ferromagnetic (FM) ordered phase with nonzero magnetization, while for $h > h_{c,\mathrm{GS}}$, it is in a paramagnetic (PM) disordered phase with zero magnetization.

Though free energy is not a well-defined observable in the MDITE process, the magnetization of spins, $\langle m\rangle \equiv \langle \sum_{i=1}^N Z_i/N\rangle$, remains accessible and physical, serving as an important diagnosis for characterizing the behaviors of the steady state $\rho_{\infty}$.

Specifically, for the TFIM, we numerically calculate the absolute magnetization $\langle |m|\rangle$, while also considering the second moment $\langle m^2\rangle$ and the fourth moment $\langle m^4\rangle$, from which the dimensionless Binder ratio $R_2 = \langle m^4\rangle / \langle m^2\rangle^2$ is obtained.
Note that $\langle |m|\rangle$ does not correspond to $\langle |\sum_{i=1}^N Z_i/N| \rangle$. Instead, in QMC simulations, $\langle |m|\rangle$ represents the ensemble average of the absolute magnetization measured in individual classical QMC samples~\cite{Sandvik2010lecture}. 
For the standard TFIM, observables $\langle |m| \rangle$ and $R_2$ are important for identifying the phase transition point from finite-size simulations~\cite{blote2002cluster,huang2020worm,sandvik2003stochastic,zhou2022quantum,yan2023quantum}. 

In the FM phase, the magnetization distribution is 
double-peaked around $\pm \langle |m| \rangle$, 
serving as a direct signature of spontaneous 
symmetry breaking (SSB). This behavior leads to $R_2 \to 1$, 
while in the PM phase the distribution becomes Gaussian, 
yielding $R_2 \to 3$. Remarkably, as we will show below, 
these observables can also be used to characterize the 
mixed-state phase transitions of $\rho_{\infty}$ in the 
presence of measurements. 

\subsection{Steady state}
\begin{figure}[htbp]
    \centering
    \subfigure[]{\includegraphics[width=0.4\textwidth]{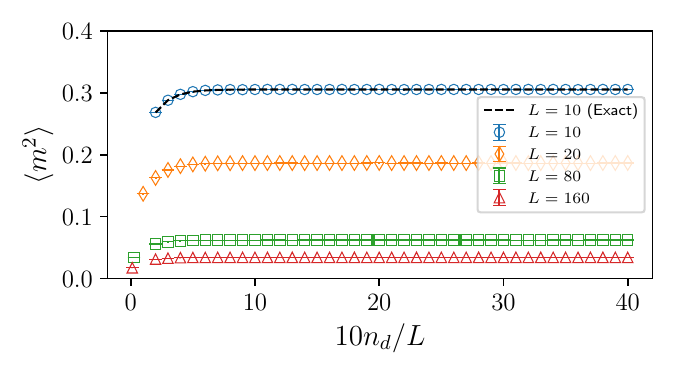}}
    \subfigure[]{\includegraphics[width=0.4\textwidth]{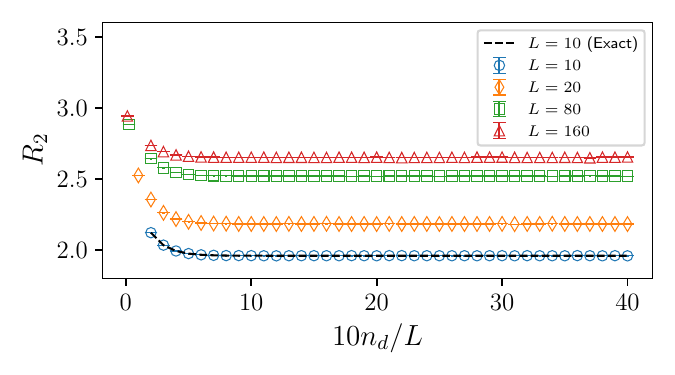}}
    \caption{
        When setting $(\tau,h,p)=(1,1.8,0.66)$ for the MDITE with the 1D TFIM:
(a) Convergence of the second moment $\langle m^2\rangle$ with increasing the number of time steps for various system sizes $L$.
(b) Convergence of the Binder ratio $R_2$ with increasing time steps for various system sizes $L$.
    }
    \label{fig:1d_convergence}
\end{figure}

We next demonstrate the existence of the steady state in the MDITE protocol.
Fig.~\ref{fig:1d_convergence} shows that for the control parameters $(\tau, h, p) = (1, 1.8, 0.66)$—which we will later identify as a critical point—the state $\rho_{n_d}$ converges to a steady state in the limit $n_d \to \infty$. 

This is evidenced by the rapid convergence of both $\langle m^2 \rangle$ and the Binder ratio $R_2$ with increasing the number of time steps, producing dynamics 
reminiscent of the quantum Zeno effect~\cite{Misra1977zeno,
Itano1990zeno,Koshino2005zeno}.
We observe similar convergence behavior for other parameter choices.

In practice, we find that $n_d=2L/\tau$ is sufficient to reach the steady state in the setup of 1D TFIM, implying that the dynamical exponent at the mixed-state critical points satisfies $z\leq 1$. In Sec.~\ref{sec:dynamical_tfim}, we perform simulations directly at the critical points and find $z\approx 1$ for all cases studied.

\subsection{Decoherence-induced phase transitions}\label{sec:mixed_mipt_numerical}
\begin{figure*}[ht!]
\centering
\begin{overpic}[width=0.32\textwidth]{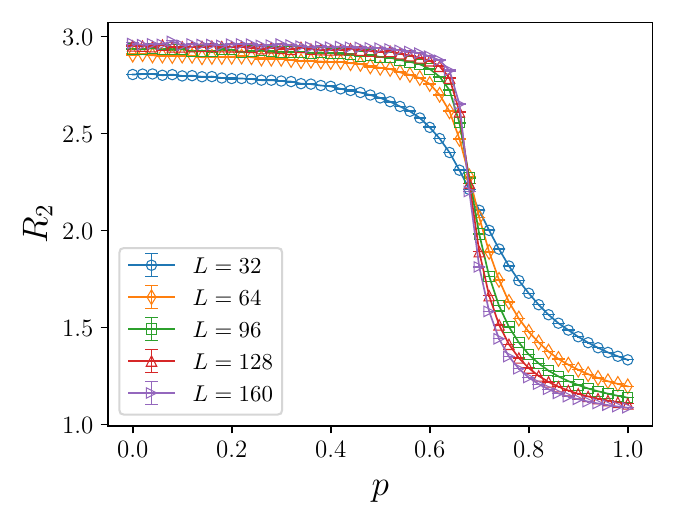}
\put (85,65) {{\textbf{(a)}}}
\end{overpic}
\begin{overpic}[width=0.32\textwidth]{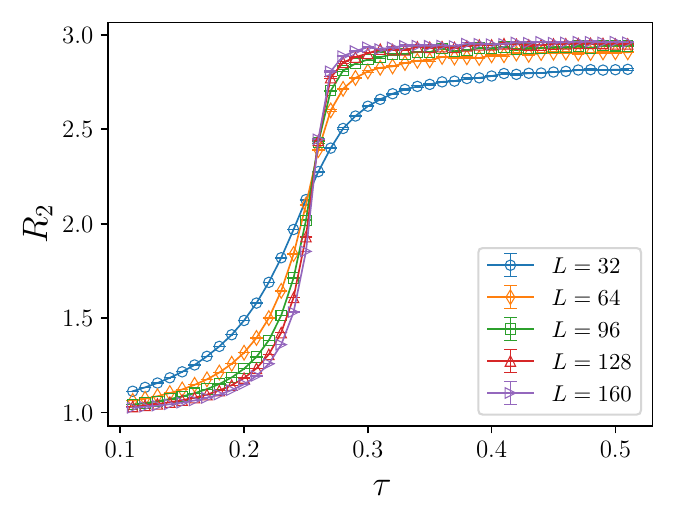}
\put (85,55) {{\textbf{(c)}}}
\end{overpic}
\begin{overpic}[width=0.32\textwidth]{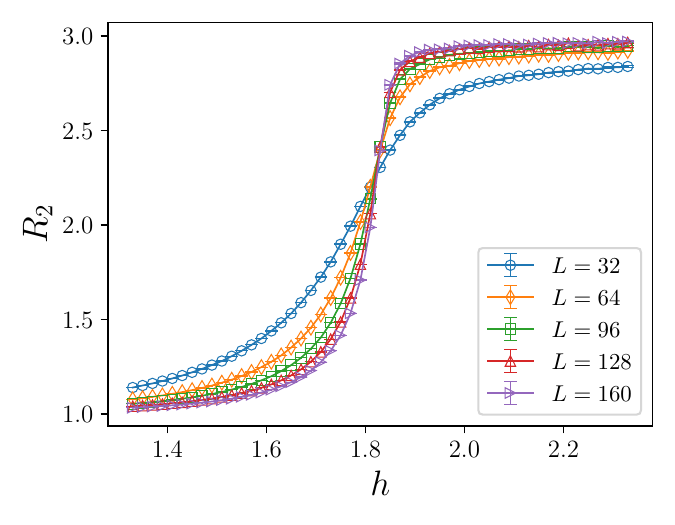}
\put (85,55) {{\textbf{(e)}}}
\end{overpic}
\begin{overpic}[width=0.32\textwidth]{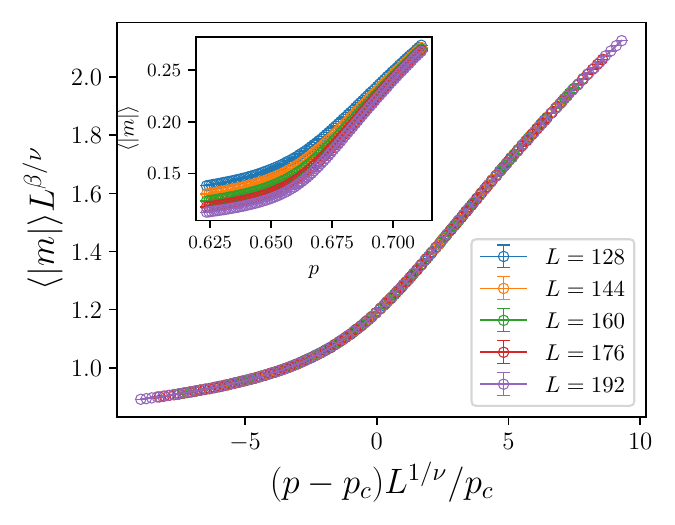}
\put (85,55) {{\textbf{(b)}}}
\end{overpic}
\begin{overpic}[width=0.32\textwidth]{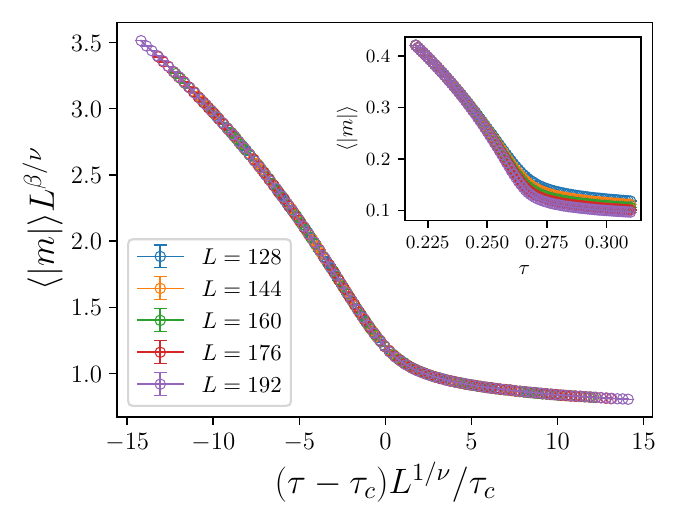}
\put (85,25) {{\textbf{(d)}}}
\end{overpic}
\begin{overpic}[width=0.32\textwidth]{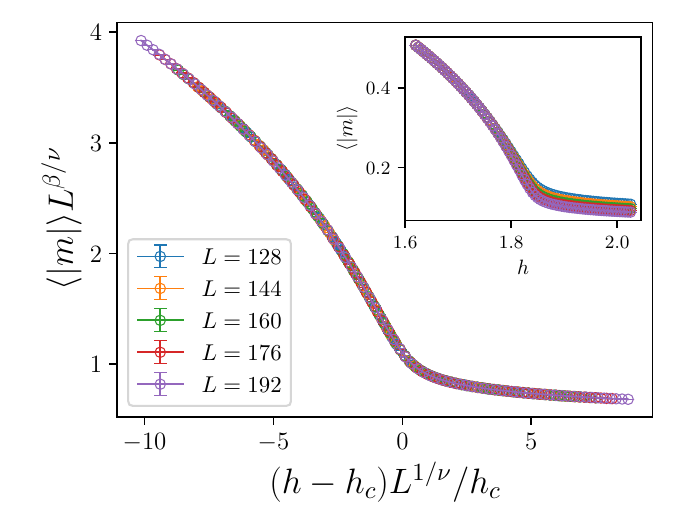}
\put (85,25) {{\textbf{(f)}}}
\end{overpic}
\caption{
For $(\tau,h)=(1,1.8)$ in the MDITE of the 1D TFIM:
(a) Binder ratios $R_2$ for different system sizes $L$ cross at $p_c \approx 0.667$;
(b) The inset panel shows $\langle |m| \rangle$ as a function of $p$ for various $L$, while the main panel presents the finite-size scaling and data collapse of $\langle |m|\rangle$. 
\\
For $(h,p)=(2.5,0.5)$:
(c) Binder ratios $R_2$ for different system sizes $L$ cross at $\tau_c \approx 0.265$;
(d) $\langle |m| \rangle$ as a function of $\tau$ for various $L$ and data collapse of $\langle |m|\rangle$.
\\
For $(\tau,p)=(1.2,0.8)$, (e) Binder ratios $R_2$ for different system sizes $L$ cross at $h_c \approx 1.84$; (f) $\langle |m| \rangle$ as a function of $h$ for various $L$ and data collapse of $\langle |m|\rangle$.
}
\label{fig:tfim_examples}
\end{figure*}

We first explore the 3D parameter space $(\tau, h, p)$ by fixing the ITE parameter $\tau$ and the Hamiltonian coupling $h$, and then scanning the measurement rate $p$ to examine whether it induces a phase transition. Our extensive numerical calculations show that tuning $p$ indeed drives a transition for a wide range of $(\tau, h)$ values when $h>1$. 

One example is shown for $(\tau, h) = (1, 1.8)$ in Fig.~\ref{fig:tfim_examples}(a), where a clear crossing of the Binder ratio $R_2$ for different system sizes $L$ identifies a critical point $p_c$. The asymptotic behavior of $R_2$—approaching $3$ for $p < p_c$ and $1$ for $p > p_c$—provides strong evidence for a mixed-state phase transition.
Moreover, since the measurement channel preserves the weak (or average) $\mathbb{Z}_2$ symmetry of the state~\cite{sala2024swssb}, the observed transition supported by the behavior of binder ratios thus corresponds to a weak-to-trivial SSB, analogous to thermal phase transitions in the 2D classical Ising model.
This separates a disordered phase for $p < p_c$ and an ordered phase for $p > p_c$.

Note that when $p = 0$, the steady state reduces to the ground state of the 1D TFIM, which is in the PM phase when $h = 1.8 > 1$. The results in Fig.~\ref{fig:tfim_examples}(a) thus indicate that sufficiently strong projective measurements can drive the system into a mixed-state FM phase, polarized along the $z$-axis. 

To extract the critical exponents of the critical point, we perform a finite-size scaling analysis \cite{privman1990finite} on 
\begin{equation}
    x_L = \frac{p-p_c}{p_c}L^{1/\nu},\quad 
    y_L = \langle |m|\rangle L^{\beta/\nu}, 
\end{equation}
where $\nu$ and $\beta$ denote the critical exponents related to the correlation length and the magnetization. 
From the data collapse, we extract $p_c\approx 0.667 $, $\nu \approx 1.08$, and $\beta \approx 0.43$, 
whose numerical stability is reported in Table~\ref{table:tfim_exponents}. 
The corresponding plots are presented in Fig.~\ref{fig:tfim_examples}(b). 

Importantly, the critical exponents $\nu$ and $\beta$ are quite different from that of the equilibrium 2D or 3D classical Ising criticality, which means this mixed-state phase transition is at least not a standard Ising one, though the numerical results indicate an Ising-like $\mathbb{Z}_2$ SSB.

\begin{table}[htbp]
\centering
\begin{tabular}{c c c c c}
\toprule
Parameters & $L_{\min}$ & Critical point & $\nu$ & $\beta$ \\
\midrule

$(\tau,h)=(1,1.8)$
& 112 & $p_c=0.6668(6)$ & 1.073(5) & 0.422(8) \\
& 128 & $p_c=0.6667(7)$ & 1.077(6) & 0.424(9) \\
& 144 & $p_c=0.6665(9)$ & 1.079(7) & 0.43(1) \\
& 160 & $p_c=0.666(1)$ & 1.084(8) & 0.43(1) \\

\midrule

$(h,p)=(2.5,0.5)$
& 112 & $\tau_c=0.2651(1)$ & 1.1873(2) & 0.480(1) \\
& 128 & $\tau_c=0.2652(1)$ & 1.189(1) & 0.480(2) \\
& 144 & $\tau_c=0.2651(2)$ & 1.190(2) & 0.480(4) \\
& 160 & $\tau_c=0.2652(4)$ & 1.191(6) & 0.480(7) \\

\midrule

$(\tau,p)=(1.2,0.8)$
& 112 & $h_c=1.8375(3)$ & 1.181(5) & 0.46(1) \\
& 128 & $h_c=1.8373(7)$ & 1.18(1) & 0.46(1) \\
& 144 & $h_c=1.837(1)$  & 1.19(2) & 0.46(1) \\
& 160 & $h_c=1.835(2)$ & 1.21(3) & 0.46(1) \\

\bottomrule
\end{tabular}

\caption{
Critical exponents obtained from data collapse
of $\langle |m|\rangle$ with $L=112,128,144,160,176,192$.
To test the stability of the extracted exponents, we first discard the $L=112$ data, setting $L_{\min}=128$, and then progressively increase $L_{\min}$ to $160$ to retain only the three largest system sizes.
}
\label{table:tfim_exponents}
\end{table}

\subsection{$\tau$- and $h$-induced phase transitions}\label{sec:tfim_tau_h_induced}
In addition to the decoherence-induced phase transitions driven by tuning $p$, the steady state exhibits mixed-state phase transitions when either the $\tau$ or $h$ is varied, with the other parameters fixed. 

As an example, for $(h,p)=(2.5,0.5)$, the Binder ratio $R_2$ shows a clear crossing at $\tau_c \approx 0.265$, with a scaling analysis of $\langle |m| \rangle$ yielding $\nu \approx 1.19$, $\beta\approx 0.48$ [Fig.~\ref{fig:tfim_examples}(c, d), Table~\ref{table:tfim_exponents}]. 

Similarly, for $(\tau,p)=(1.2,0.8)$, we identify a transition at $h_c \approx 1.84$, characterized by $\nu \approx 1.18$ and $\beta \approx 0.46$ [Fig.~\ref{fig:tfim_examples}(e, f), Table~\ref{table:tfim_exponents}].  
Notably, the values of $\nu$ and $\beta$ change for different critical points. 

\subsection{Dynamical scaling}\label{sec:dynamical_tfim}
To better understand these mixed-state phase transitions, we next extract the dynamical critical exponent $z$ by simulating the relaxation dynamics from the initial state to the steady state. 
Specifically, we determine the relaxation time $\xi_{\tau}$ by fitting the convergence of the order parameter $\langle |m_z| \rangle$ as a function of the number of time steps $n_d$ using the fitting formula 
\begin{equation}\label{eq:fit_z}
    m(n_d) = a \exp(-n_d/\xi_{\tau}) + b,
\end{equation}
for each system size. 
At the critical point, the relaxation time is expected to scale as $\xi_{\tau} \propto L^{z}$. 

\begin{figure}[htbp]
    \centering
    \subfigure[]{\includegraphics[width=0.35\textwidth]{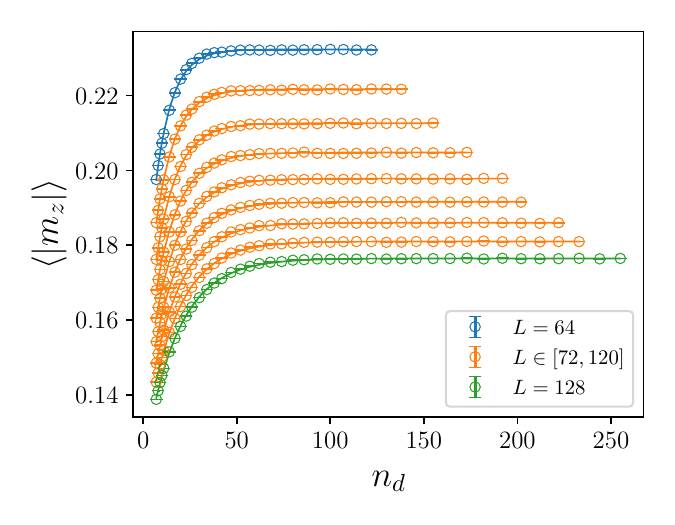}}
    \subfigure[]{\includegraphics[width=0.35\textwidth]{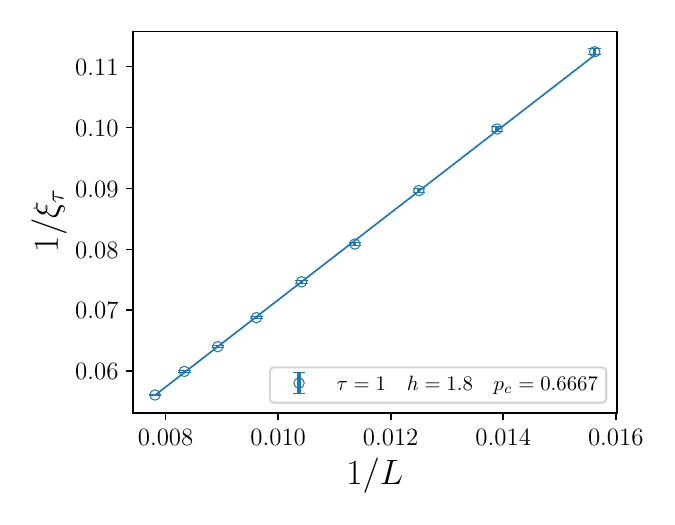}}
    \caption{
        When setting $(\tau,h,p)=(1,1.8,0.6667)$ for the MDITE with the 1D TFIM: (a)  The relaxation process of $\langle |m| \rangle$ with increasing the number of time steps $n_d$ for various system sizes $L$. It satisfies the formula $m=a\exp(-n_d/\xi_\tau)+b$, where $a$, $\xi_\tau$ and $b$ are fitting parameters.
(b) Extracting the dynamical exponent $z$ from $\xi_\tau$ with various system size $L$. The fitting formula is $\xi^{-1}_\tau=kL^{-z}$, where $k$ and $z$ are fitting parameters.
    }
    \label{fig:1d_tfim_z}
\end{figure}

For the measurement-induced critical point $(\tau,h,p)=(1,1.8,0.6667)$ in Sec.~\ref{sec:mixed_mipt_numerical},  we obtain $z \approx 1$. 
The data are shown in Fig.~\ref{fig:1d_tfim_z}, and the stability analysis yields consistent results, as summarized in Table~\ref{table:tfim_p_example_z}. 

\begin{table}[h!]
    \centering
    \begin{tabular}{c c c c c c c c}
        \toprule
        $L_{\min}$ & $k$ &$z$  \\
        \midrule
        64 &7.1(2) & 0.997(5)   \\
        72 &6.9(2) & 0.993(6)   \\ 
        80 &6.7(3) & 0.987(7)   \\ 
        88 &6.5(3) & 0.978(9)   \\ 
        96 &6.8(4) & 0.99(1)  \\ 
        \bottomrule
    \end{tabular}
    \caption{
        For $(\tau,h,p)=(1,1.8,0.6667)$, we extract the dynamical exponent $z$ from the relaxation correlation length using system sizes $L=64,72,..., 128$. To test the stability of the extracted exponent, small system sizes are gradually discarded. $L_{\min}$ denotes the smallest systems size used in each fitting.
        }
    \label{table:tfim_p_example_z}
\end{table}

Similarly, we extract the dynamical exponents for the other two parameter sets $(\tau,h,p)=(1.2,1.8373,0.8)$ and $(0.4675,2.5,0.5)$, as reported in Sec.~\ref{sec:tfim_tau_h_induced}, where the phase transitions are driven by the transverse field $h$ and the ITE parameter $\tau$, respectively. 
As reported in Appendix~\ref{appx:tfim_z}, the exponents are also close to one. 

After accounting for finite-size effects, our numerical results are consistent with a dynamical critical exponent $z\approx 1$, suggesting emergent space--time isotropy at the steady-state critical points. Although this behavior is compatible with an emergent Lorentz-invariant description, it does not by itself establish conformal symmetry. We therefore examine whether the critical points admit an effective CFT description. As demonstrated below, the numerical four-point correlation functions deviate from the cross-ratio dependence required by conformal symmetry in $(1+1)$ dimensions.

\subsection{Conformal cross-ratio function}
The squared magnetization is the spatially averaged two-point correlation function,
\begin{equation}
    \langle m^2\rangle
    \propto
    \sum_{i,j}
    \langle Z_i Z_j\rangle .
\end{equation}
At a scale-invariant critical point, the finite-size scaling forms are
\begin{equation}
    \langle |m| \rangle \sim L^{-\beta/\nu},
    \qquad
    \langle m^2\rangle \sim L^{-2\beta/\nu}.
\end{equation}
Thus, $\langle m^2\rangle$ scales with the same exponent as a two-point correlation function evaluated at a separation of order $L$.

If the critical point is described by a CFT, the microscopic operator $Z_i$ admits the coarse-grained expansion
\begin{equation}
    Z_i \sim A_0 \phi(x_i)+\cdots,
\end{equation}
where $\phi$ is the leading scaling operator with the same symmetry quantum numbers as $Z_i$. Its scaling dimension is therefore $ \Delta_\phi=\beta/\nu$.

However, the scaling of the two-point correlation function alone is insufficient to establish conformal symmetry. A stronger test is provided by the four-point correlation function. For four identical scalar primary operators $\phi$ with scaling dimension $\Delta_\phi$, conformal symmetry constrains the correlator to the form
\begin{equation}
    \left\langle
        \phi(x_1)\phi(x_2)\phi(x_3)\phi(x_4)
    \right\rangle
    =
    F(u,v)
    \prod_{i<j}
    r_{ij}^{-2\Delta_\phi/3},
\end{equation}
where $r_{ij}=|x_i-x_j|$, and $F(u,v)$ is a dimensionless function determined by the underlying CFT. Here, the conformal cross ratios are defined as 
\begin{equation}
    u=
    \frac{r_{12}r_{34}}{r_{13}r_{24}},
    \qquad
    v=
    \frac{r_{12}r_{34}}{r_{23}r_{14}}.
\end{equation}
Thus, apart from the overall power-law dependence fixed by $\Delta_\phi$, conformal symmetry restricts the coordinate dependence of the four-point correlation function to the cross ratios $u$ and $v$.

Under the conformal map $S^1\times\mathbb{R}\to\mathbb{C}$, equal-time separations between scaling operators are replaced by the chord distances
\begin{equation}\label{eq:chord-distance}
    d_{ij}
    =
    \frac{L}{\pi}
    \left|
        \sin\left[\frac{\pi(x_i-x_j)}{L}\right]
    \right|,
\end{equation}
where $L$ is the circumference of the spatial circle. 

Moreover, for four cyclically ordered points on the circle, Ptolemy's theorem yields
$v=u/(1-u)$. Thus, for equal-time four-point functions in a $(1+1)$D CFT, only one independent real cross ratio remains. Henceforth, we denote it by $\eta\equiv u$ and replace $F(u,v)$ with $F(\eta)$.

We therefore define the lattice four-point function
\begin{equation}
    F_L(\eta;\Delta_\phi)
    =
    \left(
        \prod_{i<j} d_{ij}^{\,2\Delta_\phi/3}
    \right)
    \left\langle
        Z_{x_1}Z_{x_2}Z_{x_3}Z_{x_4}
    \right\rangle_L.
\end{equation}
If the critical point is described by a CFT, then
\begin{equation}
    F_L(\eta;\Delta_\phi)
    \longrightarrow
    A_0^4 F(\eta)
\end{equation}
in the scaling limit, up to finite-size and irrelevant-operator corrections. Consequently, data obtained from different system sizes and operator configurations with the same cross ratios are expected to collapse onto a common function.

To eliminate the nonuniversal normalization associated with $A_0$ and reduce finite-size effects, we also consider the normalized lattice function
\begin{equation}
    \widetilde{F}_L(\eta;\Delta_\phi)
    =
    \frac{
        F_L(\eta;\Delta_\phi)
    }{
        F_L(\eta_0;\Delta_\phi)
    }.
\end{equation}
Here, $\eta_0$ is an arbitrary fixed reference cross ratio for which the correlator is nonzero; throughout this work, we choose $\eta_0=1/2$. In the scaling limit,
\begin{equation}
    \widetilde{F}_L(\eta;\Delta_\phi)
    \longrightarrow
    \frac{F(\eta)}{F(\eta_0)},
\end{equation}
so that the nonuniversal factor $A_0^4$ cancels.

A convenient one-parameter family of four points is
\begin{equation}
    x_1=0,\quad
    x_2=\ell,\quad
    x_3=\frac{L}{2},\quad
    x_4=\frac{L}{2}+\ell,
\end{equation}
where $L$ is chosen to be an integer multiple of $16$, and
\begin{equation}
    \ell\equiv \ell(n)=\frac{nL}{16},
    \qquad
    n=2,3,4,5,6.
\end{equation}
We exclude $n=1$ and $n=7$ to suppress short-distance lattice effects and avoid configurations with nearly coincident points. The reference value $\eta_0=1/2$ corresponds to $n=4$, for which $\ell=L/4$.

\begin{figure}[htbp]
    \centering
    \subfigure[]{\includegraphics[width=0.35\textwidth]{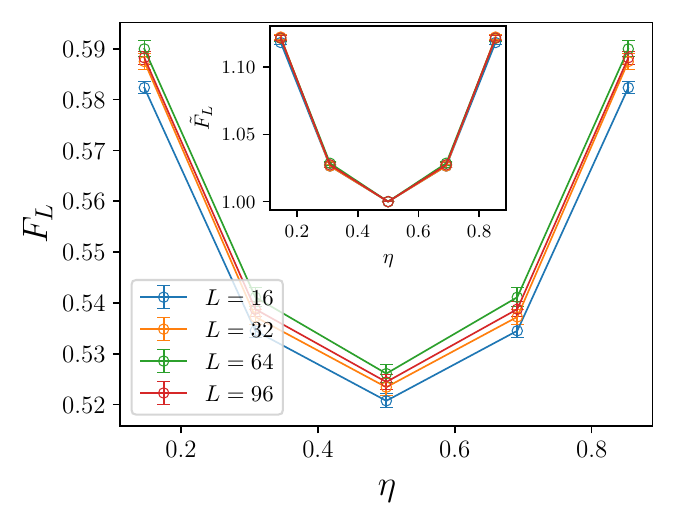}}
    \subfigure[]{\includegraphics[width=0.35\textwidth]{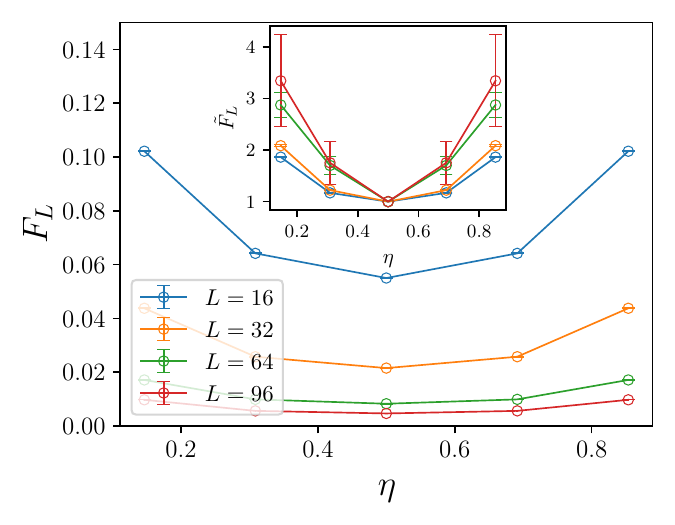}}
    \caption{
       Four-point cross-ratio functions for
        (a) the conventional $(1+1)$D TFIM at its
        ground-state critical point and
        (b) the critical steady state of the MDITE model at $(\tau,h,p)=(1,1.8,0.6667)$.
        The main panels show the unnormalized lattice function
        $F_L(\eta;\Delta_\phi)$, whereas the insets show the normalized
        function $\widetilde{F}_L(\eta;\Delta_\phi)$, with normalization point
        $\eta_0=1/2$. The TFIM data exhibit an excellent collapse across
        different system sizes, consistent with the Ising CFT. 
        By contrast, the MDITE data exhibit a less satisfactory collapse, with
        noticeable residual system-size dependence. This behavior suggests possible
        deviations from conformal invariance.
    }
    \label{fig:1d_four_point}
\end{figure}

As an example, Fig.~\ref{fig:1d_four_point} compares the unnormalized lattice four-point
function $F_L(\eta;\Delta_\phi)$ and its normalized version
$\widetilde{F}_L(\eta;\Delta_\phi)$ for the conventional $(1+1)$D TFIM
at its ground-state critical point [Fig.~\ref{fig:1d_four_point}(a)] and for
the critical steady state at $(\tau,h,p)=(1,1.8,0.6667)$ of the MDITE model
[Fig.~\ref{fig:1d_four_point}(b)].
For the TFIM, data from different system sizes exhibit an excellent collapse
onto a common function of $\eta$, consistent with the conformal invariance of
the $2$D Ising critical point. By contrast, the MDITE data show a less
satisfactory collapse, with noticeable residual system-size dependence. This
behavior points to possible deviations from conformal invariance at the
finite-parameter MDITE critical point.

\section{Analysis of Criticality in MDITE Steady States}
\label{sec:theoretical}

To understand the mechanism underlying the mixed-state phase transitions observed in the numerical simulations of the previous section, we consider the continuous-parameter limit $\tau \to 0$, $ p \to 0$ and $ \tau \sim p$.
This limit provides a controlled setting in which imaginary-time filtering and measurement-induced decoherence compete on an equal footing. 
As we show below, this limit reproduces the qualitative phase structure observed in the finite-parameter simulations.
We emphasize, however, that the continuous-parameter limit should be regarded as a theoretically controlled benchmark for understanding the transition mechanism, rather than as a quantitatively accurate description of the finite-parameter regime, for which a complete analytical treatment remains considerably more challenging.

\subsection{Choi representation}
We first represent the steady state in the double-Hilbert space formalism.
Using the identities
\begin{equation}
    I_i=\ket{0_i}\bra{0_i} + \ket{1_i}\bra{1_i}, \quad
    Z_i = \ket{0_i}\bra{0_i} - \ket{1_i}\bra{1_i},
\end{equation}
the local measurement channel in Eq.~\eqref{eq:local_channle} can be rewritten as 
\begin{equation}
    \mathcal{E}_{p,i}(\rho) = \bigg(1-\frac{p}{2}\bigg) \rho + \frac{p}{2} Z_i \rho Z_i. 
\end{equation}

Under the Choi-Jamiołkowski isomorphism~\cite{Choi1975,Jamiolkowski1972}, $\mathcal{E}_{p,i}$ can be mapped onto a superoperator acting on a doubled Hilbert space,  which is 
\begin{equation}
    \mathcal{E}_{p,i} \longrightarrow 
    \hat{\mathcal{E}}_{p,i}
    = (1-p)^{1/2} e^{\tilde{\tau}\, Z_i^{L} Z_i^{R}},
\end{equation}
where the inter-copy coupling strength is given by
\begin{equation}
    \tilde{\tau} \equiv \tilde{\tau}(p)
    = \tanh^{-1}\!\left( \frac{p}{2-p} \right),
\end{equation}
and superscripts $L$ and $R$ denote operators acting on the left and right copies of the Hilbert space, respectively. In addition, for $x\equiv p/(2-p) \in [0, 1]$, we can rewrite 
 \begin{equation}
    \tilde\tau = \tanh^{-1}(x)
    = \frac{1}{2}\ln\bigg( \frac{1+x}{1-x} \bigg) 
    = 
    -\frac{1}{2}\ln(1-p).
 \end{equation}
When $\tilde{\tau}\to 0$, $\tilde{\tau}$ can be approximately replaced by $p/2$.

Accordingly, the evolution described in Sec.~\ref{sec:protocol} becomes 
\begin{equation}\label{eq:evo_sp}
    \dket{\rho_k}
    \propto
    e^{-\tau \tilde{H}}
    e^{\tilde{\tau} \sum_i Z_i^{L} Z_i^{R}}
    \dket{\rho_{k-1}},
\end{equation}
with the two-copy Hamiltonian defined as
\begin{equation}
    \tilde{H} \equiv \frac{1}{2} (H^{L} + H^{R} ).
\end{equation}

\subsection{Effective Hamiltonian in the continuous limit}
In the limit $\tau, \tilde{\tau} \to 0$, with $\tau \sim \tilde{\tau}$, 
Eq.~\eqref{eq:evo_sp} can be simplified using the Baker-Campbell-Hausdorff (BCH) formula. 
Keeping terms up to linear order in $\tau$ and neglecting $O(\tau^2)$ corrections, we obtain
\begin{equation}
    e^{-\tau \tilde{H}}
    e^{\tilde{\tau} \sum_i Z_i^{L} Z_i^{R}}
    \approx
    e^{-\tau H_{\text{eff}}},
\end{equation}
where the effective Hamiltonian is given by
\begin{equation}\label{eq:heff_combined}
   H_{\text{eff}}
   \equiv
   \tilde{H}
   -
   \frac{p}{2\tau}
   \sum_i Z_i^L Z_i^R.
\end{equation}

Therefore, the ITE and measurement operations can be combined into a single effective transformation in this limit, which is exactly an doubled-space ITE with imaginary-time parameter $\tau$, under an effective two-copy Hamiltonian $H_{\mathrm{eff}}$. 

Moreover, the initial state in the doubled space is given by
\begin{equation}
   \rho_0 \mapsto \dket{\rho_0} \propto \sum_{\alpha}\ket{\alpha}\otimes \ket{\alpha^*}, 
\end{equation}
where $\{\ket{\alpha}\}$ denotes an arbitrary orthonormal basis of the single-copy Hilbert space. This state is basis independent and corresponds to the maximally entangled state in the doubled space~\cite{Maruocheng2025sptmixed,zack2025renyi1corr}.
In the limit of infinite time steps ($n_d\to \infty$), the convergence of the two-copy ITE to the ground state depends on a nonzero overlap between the initial state $\dket{\rho_0}$ and the ground state of the effective Hamiltonian $H_{\mathrm{eff}}$. Such an overlap is generically expected in the absence of symmetry obstructions. Under this assumption, the steady-state problem of the MDITE in the continuous limit reduces to the ground-state problem of the doubled-space Hamiltonian $H_{\mathrm{eff}}$. 

For the example of the 1D TFIM \eqref{eq:h_tfim}, the effective Hamiltonian takes the form
\begin{equation}\label{eq:heff_tfim}
   \begin{aligned}
     {H}_{\text{eff}}=& -\frac{1}{2}\sum_{\langle ij\rangle}(Z_i^LZ_j^L+Z_i^RZ_j^R) -\frac{ h}{2}\sum_i(X_i^L + X_i^R)
     \\ 
    & -
   \frac{p}{2\tau}
   \sum_i Z_i^L Z_i^R.
   \end{aligned}
\end{equation}

When $p/\tau = 0$, the two copies are decoupled and the $H_{\text{eff}}$ has a 
$\mathbb{Z}_2^L \times \mathbb{Z}_2^R$ symmetry, where 
$\mathbb{Z}_2^{L(R)}$ corresponds to independent spin-flip symmetry in the left (right) copy, generated by
$X^{L(R)} \equiv \prod_i X_i^{L(R)}$.

\begin{figure}[htbp]
    \centering
    \subfigure[]{\includegraphics[width=0.2\textwidth]{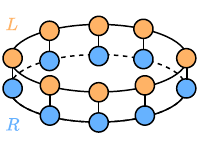}}
    \subfigure[]{\includegraphics[width=0.35\textwidth]{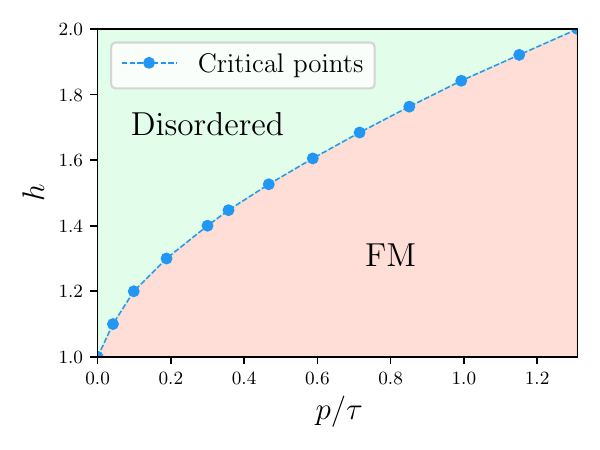}}
        \caption{
(a) Ladder geometry of $H_{\text{eff}}$ in Eq.~\eqref{eq:heff_tfim} 
for the case where $H$ is the 1D TFIM. 
The two legs correspond to the left ($L$) and right ($R$) copies of the chain 
(two rings under periodic boundary conditions). 
The interlayer Ising couplings along the rungs have strength $p/2\tau$. 
(b) Ground-state phase diagram of $H_{\text{eff}}$ obtained from 
QMC simulations.
}
    \label{fig:ladder}
\end{figure}

In the continuous-parameter limit $\tau \sim \tilde{\tau} \sim p$, the inter-copy coupling is finite, and only the diagonal spin-flip transformation preserves the Hamiltonian. 
The symmetry is therefore reduced to the diagonal subgroup $\mathbb{Z}_2^{\text{diag}}=\{1, X^L X^R\}\cong \mathbb{Z}_2$, and the effective Hamiltonian \eqref{eq:heff_tfim} is equivalent to a standard TFIM defined on a two-leg ladder geometry, as illustrated in Fig.~\ref{fig:ladder}(a).

Fig.~\ref{fig:ladder}(b) shows the ground-state phase diagram of 
$H_{\text{eff}}$ in Eq.~\eqref{eq:heff_tfim}. The critical point at $h=1$ lies at $p/\tau=0$. 
Note that the continuous limit requires $p/\tau \sim O(1)$, and this corresponds 
to the region $h>1$. Thus, sufficiently large $p/\tau$ can drive a transition 
from the disordered phase to the ordered FM phase when $h>1$ in the doubled space.

\subsection{Phase transitions in the continuous-parameter limit}
Note that in the effective Hamiltonian~\eqref{eq:heff_tfim}, both the interlayer 
Ising coupling (from measurements) and the intralayer Ising 
coupling (from the Ising interaction in the original 
Hamiltonian) favor FM ordering in the same diagonal 
$\mathbb{Z}_2$ channel, while the transverse field $h$ enhances quantum 
fluctuations and tends to disorder the system.

The key to understanding the decoherence-induced phase transitions in the 
continuous limit is the dual role of the $Z$-basis: it is both the basis 
in which projective measurements are performed and the basis in which 
the ordered phase of the Hamiltonian is defined. When $p/\tau$ is 
sufficiently large compared to $h$—equivalently, when the decoherence 
strength $p$ dominates over the coherence scale $\tau h$ via the ITE operator $e^{-\frac{\tau}{2}H}$—the system 
develops FM long-range order for the steady state in the doubled space. 

This ordered phase is characterized by the local order parameter $\langle\!\langle \rho_{\infty} | Z_i^{L(R)} | \rho_{\infty} \rangle\!\rangle$ (or one can equivalently consider the magnetization averaged over all sites),
where $\dket{\rho_{\infty}}$ denotes the ground state of 
$H_{\text{eff}}$. In the original Hilbert space, the Choi mapping relates 
this quantity to the nonlinear quantity $\mathrm{Tr}(\rho_{\infty}^2 Z_i)$,
with $\rho_{\infty}$ denoting the steady state of the MDITE.
Though the parameter region where 
$\mathrm{Tr}(\rho_{\infty}^2 Z_i)\neq 0$ needs not coincide exactly with 
that where the linear order parameter $\mathrm{Tr}(\rho_{\infty} Z_i)$ is 
nonzero, both capture the SSB physics. The resulting 
decoherence-induced phase transitions for the steady state of the MDITE protocol in the continuous-parameter limit is therefore expected to be 
qualitatively identical and within the same 2D Ising universality class.

We emphasize that the results in the continuous-parameter limit are 
qualitatively consistent with the numerical results in 
Sec.~\ref{sec:mixed_mipt_numerical} away from the continuous limits.
The mechanism 
of these mixed-state phase transitions in 
Sec.~\ref{sec:mixed_mipt_numerical} could therefore be understood as 
follows.

\subsection{Phase transitions away from the continuous limits}
In the regime $h>1$ with relatively short imaginary-time length 
($\tau h \lesssim p$), frequent projective measurements at large $p$ 
interrupt the buildup of quantum coherence.
The transverse field, which generates quantum fluctuations in the 
$X$ basis, is repeatedly suppressed before correlations can develop, 
as the state is frequently projected onto the $Z$ basis. With 
off-diagonal coherence effectively suppressed, the dynamics become 
dominated by the diagonal Ising terms, steering the system toward a 
FM phase analogous to that of the classical Ising model. 

In contrast, when the transverse field is sufficiently strong or the 
evolution time is sufficiently long ($\tau h \gtrsim p$), quantum 
coherence develops rapidly within each imaginary-time interval. 
Coherence is established faster than it can be suppressed by 
measurements, preventing the formation of FM order even at 
$p=1$. Consequently, no phase transition occurs in this regime upon 
tuning the measurement rate $p$. 
One may also expect that when $h \le 1$, no phase transition characterized by spin magnetization occurs upon tuning $p$, which has also been observed in our numerical observations.

Since the competition between quantum coherence 
and measurement-induced decoherence can establish a critical balance,
tuning the system parameters, including $p$, $\tau$, and $h$, toward 
this balance drives a mixed-state phase transition separating the 
ordered and disordered phases, as shown in Sec.~\ref{sec:tfim_tau_h_induced}
In this sense, transitions induced by 
varying $p$, $\tau$, or $h$ originate from the same underlying 
mechanism. The difference lies only in which parameter controls the 
competition.

Furthermore, while the continuous-parameter limit admits a description in terms of the effective Hamiltonian $H_{\mathrm{eff}}$ and shows critical behavior consistent
with the Ising universality class, the intermediate finite-parameter regime
appears to display different critical behavior. In this regime, away from the
controlled limiting cases, the extracted critical exponents differ
substantially from the Ising values and show an apparent dependence on the
protocol parameters $h$, $\tau$, and $p$. These results suggest that the
finite-parameter MDITE transitions may contain a richer critical structure than
that captured by the continuous-parameter limit equilibrium description.

Physically, the finite-parameter regime need not be governed by the simple local
equilibrium description obtained in the continuous limit. Formally, the two
noncommuting steps can be combined into an effective generator through the
BCH expansion,
\begin{equation}
    e^{-\tau \tilde{H}}
    e^{\tilde{\tau} \sum_i Z_i^{L} Z_i^{R}}
    \equiv 
    e^{-\tau H_{\mathrm{eff}}}.
\end{equation}
Away from the continuous limit, however, $H_{\mathrm{eff}}$ contains an
infinite series of nested commutators, which can generate complicated and
potentially nonlocal interactions in the doubled Hilbert space. This suggests
that the finite-parameter critical behavior need not coincide with the
conventional Ising description of the continuous limit, although its full
analytical classification remains open.

In addition, the emergence of long-range order with an increasing measurement rate can also be understood intuitively from the perspective of QMC simulations: a higher measurement rate naturally generates larger clusters that connect a broader range of sites in each QMC update. This provides a direct and intuitive picture of how measurements drive ordering in the system of TFIM, highlighting the use of the diagrammatic representations in Sec.~\ref{appx:pathint} and the advantage of QMC in revealing the underlying mechanism. Further details are provided in Appendix.~\ref{appx:qmc3}.

\begin{figure}[htbp]
    \centering
    \subfigure[]{\includegraphics[width=0.32\textwidth]{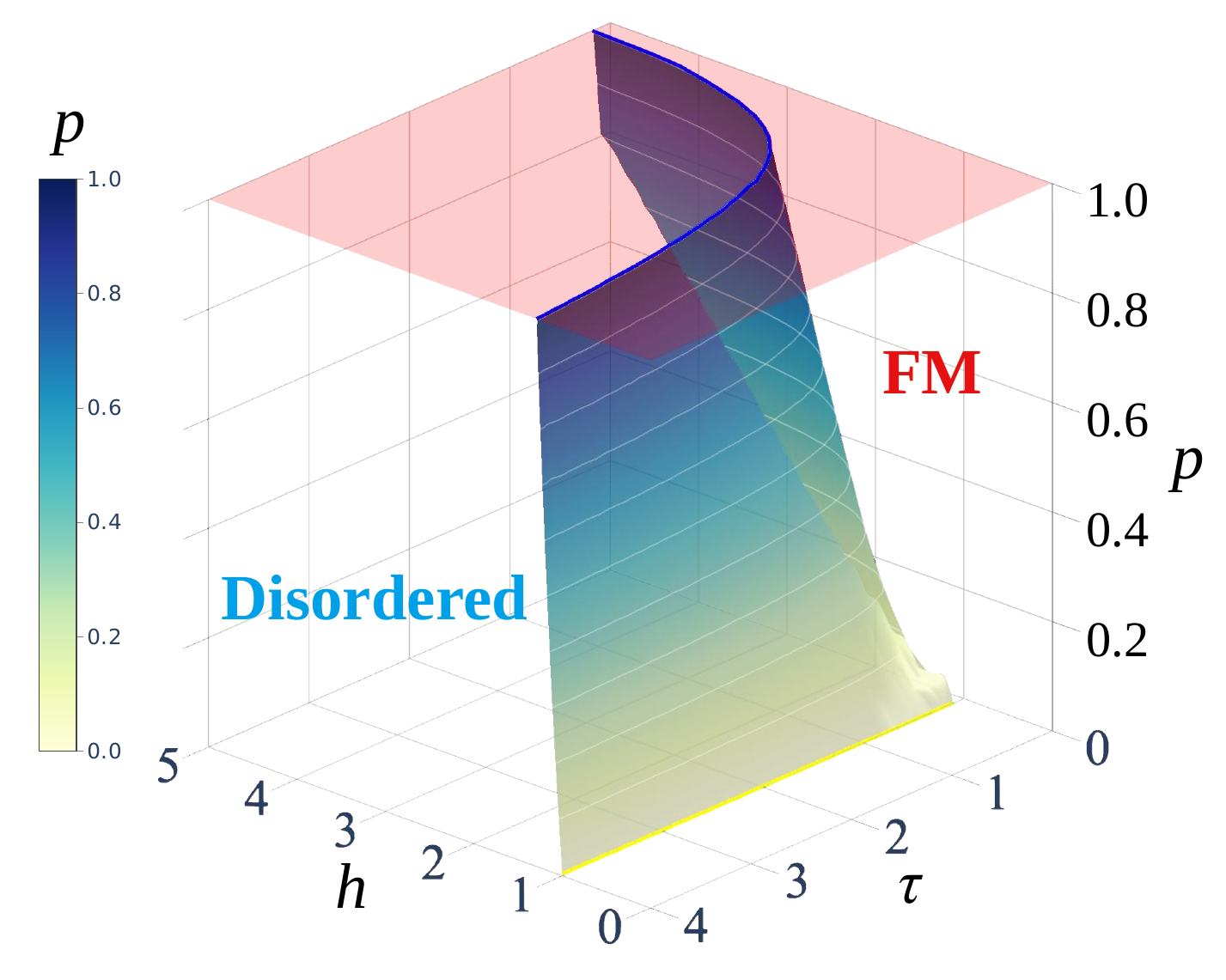}}
    \subfigure[]{\includegraphics[width=0.32\textwidth]{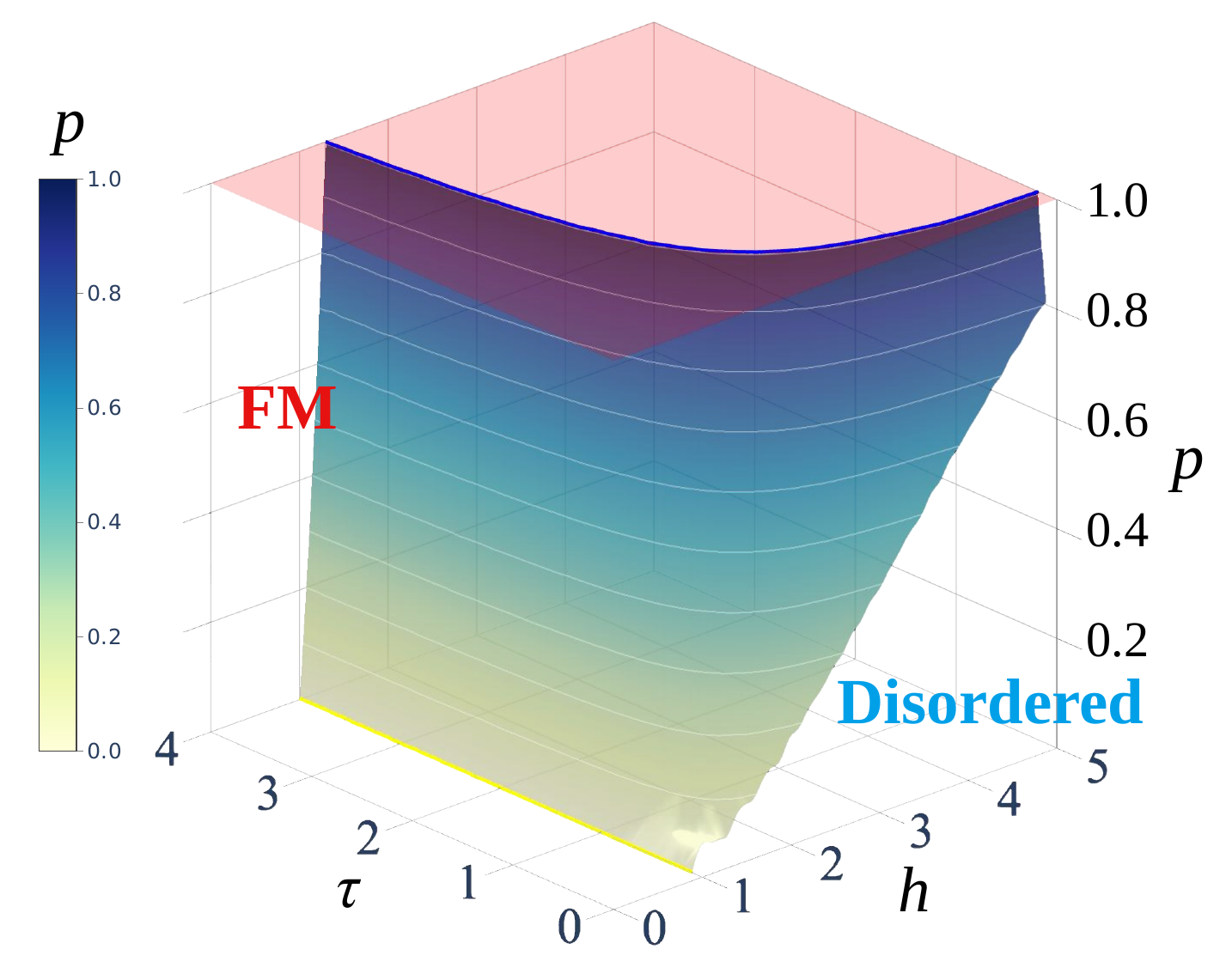}}
    \caption{
        Away from the continuous limits, critical surfaces of the steady state of the MDITE protocol viewed from different angles [(a) and (b)]. To facilitate the distinction of critical $p$ corresponding to different parameters on the critical surface, the figure employs a colorbar to map $p$ to different colors.
    }
    \label{fig:critical_surface_tfim}
\end{figure}

\subsection{Schematic transition surface in the finite-parameter regime}
To visualize how the MDITE transition evolves with the protocol parameters, we
construct a schematic transition surface from approximate transition points
estimated by Binder-ratio crossings for several system sizes, as shown in
Fig.~\ref{fig:critical_surface_tfim}. This surface should be understood as a
qualitative guide to the finite-parameter phase boundary, rather than a complete
or high-precision determination of the full critical manifold.

Within the parameter range accessible to our simulations,
Fig.~\ref{fig:critical_surface_tfim}(b) shows that the transition shifts toward
larger $\tau$ as $h$ approaches $1$. 
This trend is expected: longer evolution times require weaker transverse fields to allow FM order to emerge under strong measurements. Consequently, in the limit $h \to 1$, the critical surface is expected to extend infinitely along the $\tau$-axis, permitting a transition at any $\tau$ by tuning $p$. A similar reasoning applies in the opposite limit, $h \to \infty$ and $\tau \to 0$.

Fig.~\ref{fig:critical_surface_tfim} provides a schematic summary of the
transition points obtained in our finite-parameter simulations. This regime
should be distinguished from the continuous limit discussed in
Sec.~\ref{sec:theoretical}, where the critical behavior reduces to the Ising
universality class. Here the surface is used to illustrate how the transition
evolves with the protocol parameters in the finite-parameter regime. Although
the same QMC formulation can in principle approach the continuous limit,
resolving that regime requires $\mathcal{O}(L/\tau)$ replica states in the QMC
simulation. Therefore, as $\tau \to 0$, though the QMC update remains polynomial, the
practical computational cost increases substantially.

\section{2D model with continuous symmetry}\label{sec:result:cdhm}
\subsection{Model and observables}
\begin{figure}[ht]
    \centering
    \includegraphics[width=0.23\textwidth]{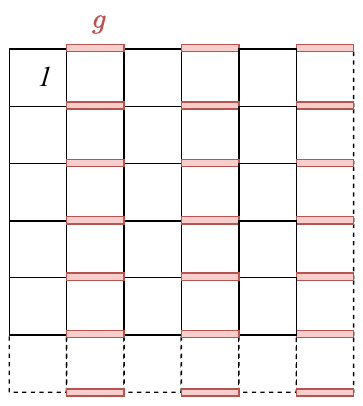}
    \caption{
        Square lattice of the 2D CDHM, where the red thick bonds denote couplings with strength $g$, while the remaining bonds have unit strength.
    }
    \label{fig:2d_lattice}
\end{figure}

As the second example, we consider the 2D $S=1/2$ columnar dimerized Heisenberg model (CDHM) on a $L\times L$ square lattice~\cite{Matsumoto2001qmcdhm}. The Hamiltonian is given by
\begin{equation}
    H=\sum_{\langle ij\rangle_1}\mathbf{S}_i\cdot\mathbf{S}_j+g\sum_{\langle ij\rangle_g}\mathbf{S}_i\cdot \mathbf{S}_j 
\end{equation}
where $\mathbf{S}_i$ is the spin operator at site $i$, and $\langle ij\rangle_1$ and $\langle ij\rangle_g$ denote two different sets of nearest-neighbor pairs, as shown in Fig.~\ref{fig:2d_lattice}.
We also consider the local $Z$-basis as the computational basis, and the periodic boundary condition is chosen.
The model has a global $SU(2)$ symmetry and the ground state exhibits a quantum phase transition at $g_{c,\mathrm{GS}}\approx 1.90951$~\cite{Sandvik2010lecture,Matsumoto2001qmcdhm,Ma2018Anomalous}, which belongs to the 3D $O(3)$ universality class. 
The critical point separates a Néel ordered phase ($g<g_{c,\mathrm{GS}}$) from a dimerized phase ($g>g_{c,\mathrm{GS}}$).

Due to the antiferromagnetic (AFM) nature of the Heisenberg interactions, we consider the staggered magnetization $\langle m_z\rangle = \Big\langle \sum_i (-1)^{x_i + y_i} Z_i / N \Big\rangle$ as the order parameter, where $x_i$ and $y_i$ denote the integer coordinates of site $i$ in the lattice. As in the TFIM, we compute the absolute magnetization $\langle |m_z| \rangle$, the second and fourth moments of the magnetization, and the Binder ratio $R_{2,z}$. 
Furthermore, for the ground state of the CDHM, the Binder ratio associated with the $O(3)$ phase transition approaches $R_{2,z} \to 9/5$ in the Néel phase and $R_{2,z} \to 3$ in the dimerized phase~\cite{sandvik1997finite,wang2006high,ding2018engineering,Ma2018Anomalous}.

\subsection{Steady state}
\begin{figure}[htbp]
    \centering
    \subfigure[]{\includegraphics[width=0.4\textwidth]{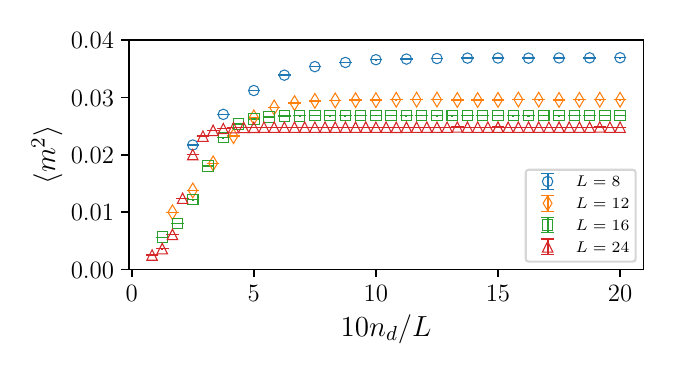}}
    \subfigure[]{\includegraphics[width=0.4\textwidth]{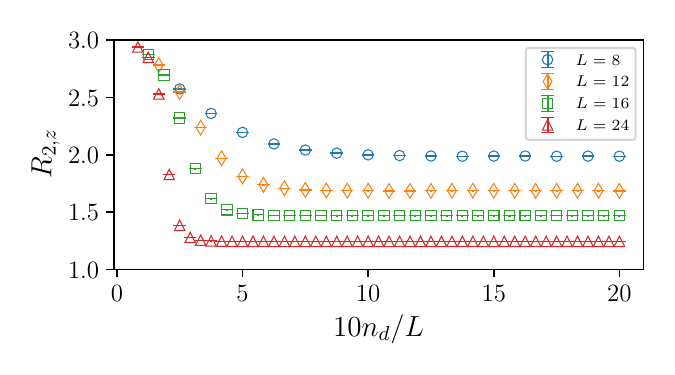}}
    \caption{
        When setting $(\tau,g,p)=(1,3.5,0.5)$ for the MDITE with the 2D CDHM: (a) Convergence of the squared magnetization $\langle m^2 \rangle$ with increasing the number of time steps $n_d$ for various system sizes $L$. 
(b) Convergence of the Binder ratio $R_{2,z}$ with increasing $n_d$ for various system sizes $L$.
    }
    \label{fig:2d_convergence}
\end{figure}

As in the 1D TFIM example, the MDITE protocol in this case also drives the system toward a steady state in the limit of large number of time steps $n_d$. Fig.~\ref{fig:2d_convergence} illustrates this convergence for representative parameters $(\tau, g, p)=(1,3.5,0.5)$, where both the staggered magnetization $\langle |m_z| \rangle$ and the Binder ratio $R_{2,z}$ quickly approach their stationary values. In practice, we again find that $n_d = 2L/\tau$ is sufficient to reach the steady state, and this choice is used in all subsequent QMC simulations for the CDHM. 
As discussed later, the dynamical exponent $z$ for the mixed-state phase transitions in this example is also around one. 

\subsection{Mixed-state phase transitions}

\begin{figure*}[ht!]
\centering
\begin{overpic}[width=0.32\textwidth]{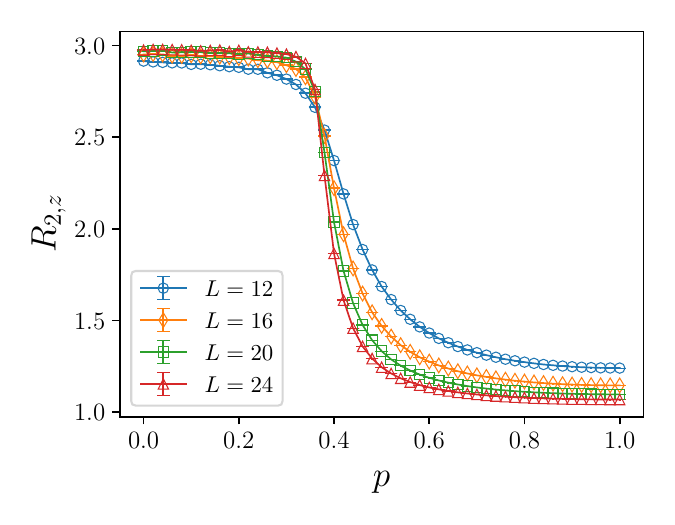}
\put (82,65) {{\textbf{(a)}}}
\end{overpic}
\begin{overpic}[width=0.32\textwidth]{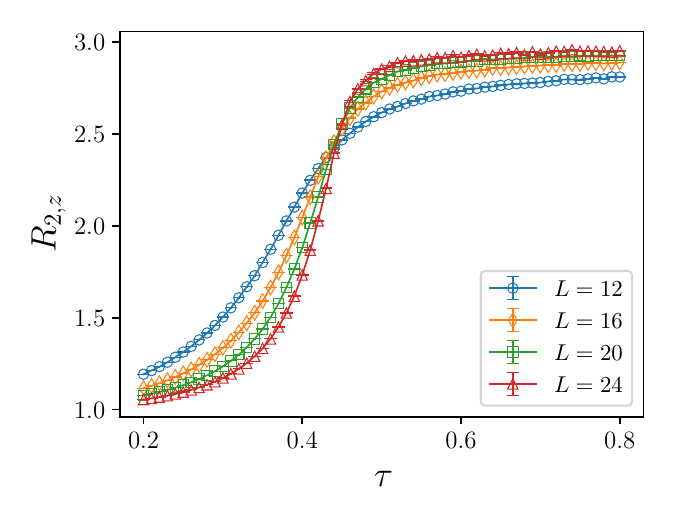}
\put (22,65) {{\textbf{(c)}}}
\end{overpic}
\begin{overpic}[width=0.32\textwidth]{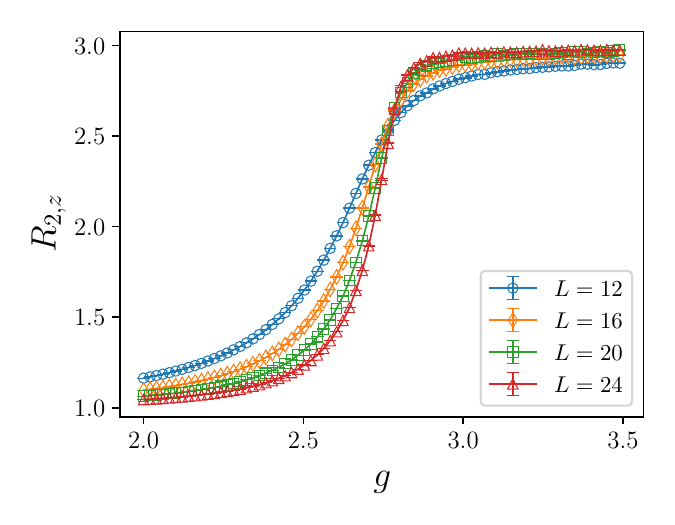}
\put (22,65) {{\textbf{(e)}}}
\end{overpic}
\begin{overpic}[width=0.32\textwidth]{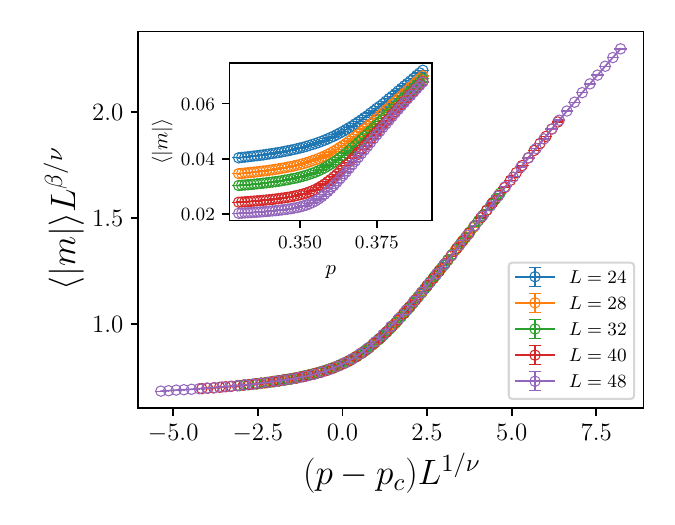}
\put (25,25) {{\textbf{(b)}}}
\end{overpic}
\begin{overpic}[width=0.32\textwidth]{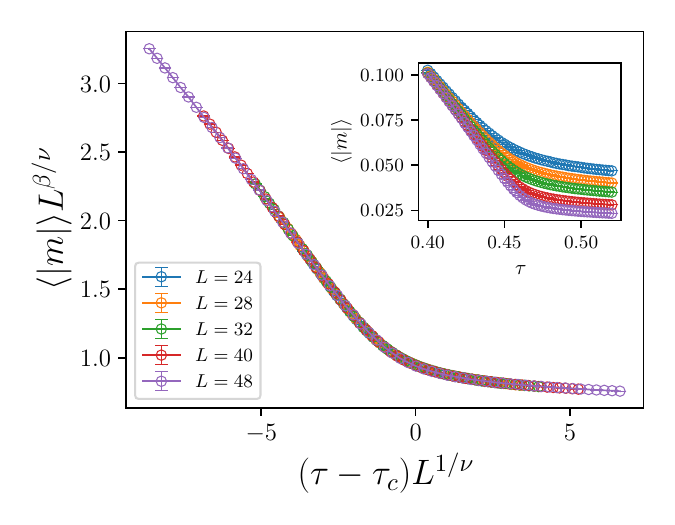}
\put (83,25) {{\textbf{(d)}}}
\end{overpic}
\begin{overpic}[width=0.32\textwidth]{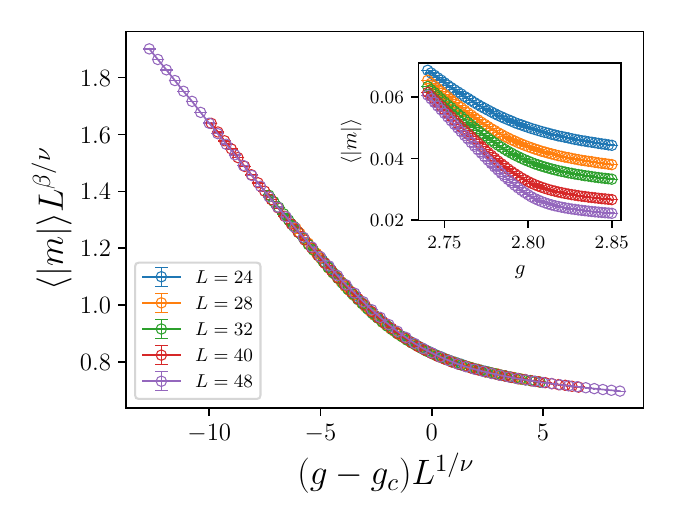}
\put (83,25) {{\textbf{(f)}}}
\end{overpic}
\begin{overpic}[width=0.32\textwidth]{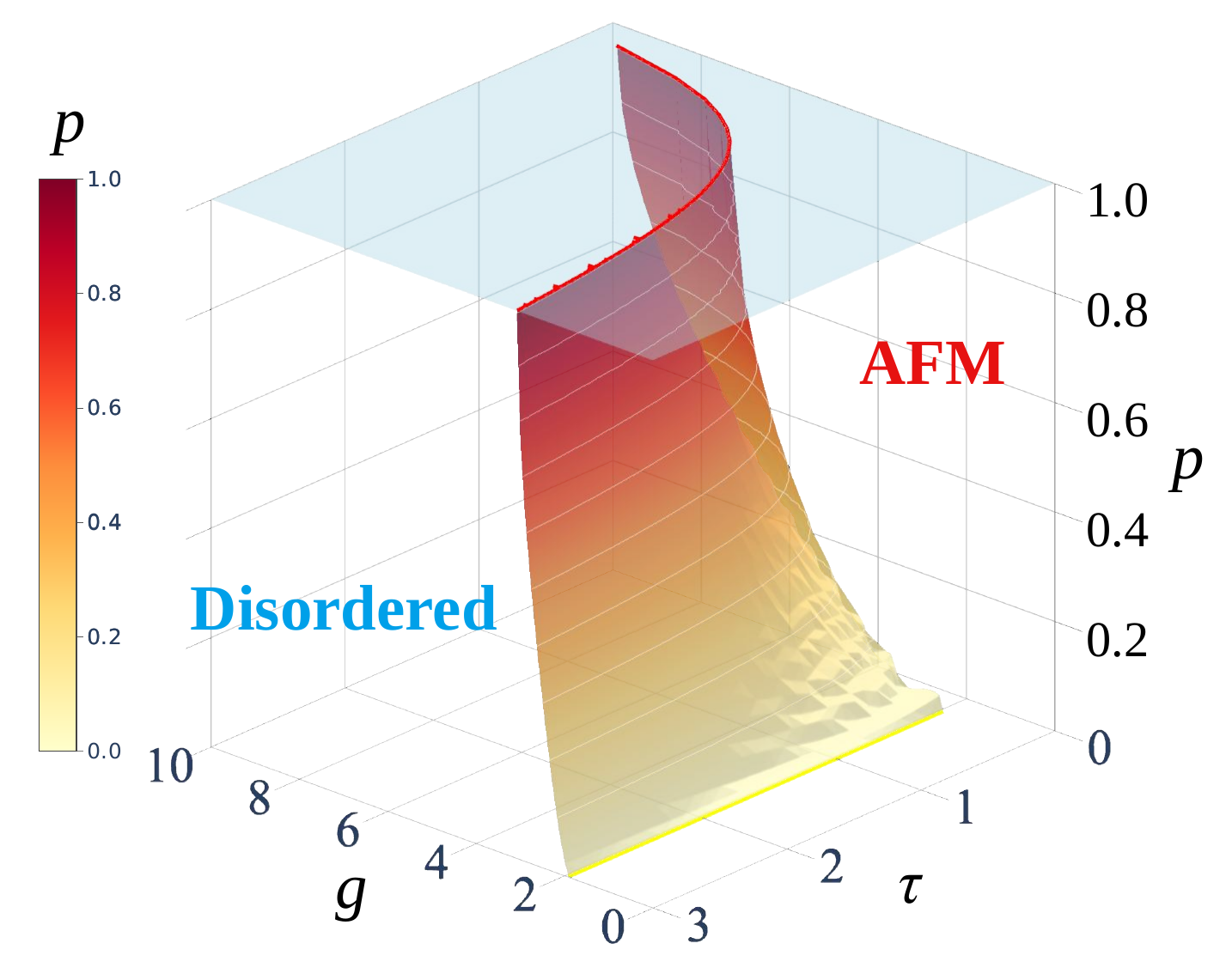}
\put (5,5) {{\textbf{(g)}}}
\end{overpic}
\begin{overpic}[width=0.32\textwidth]{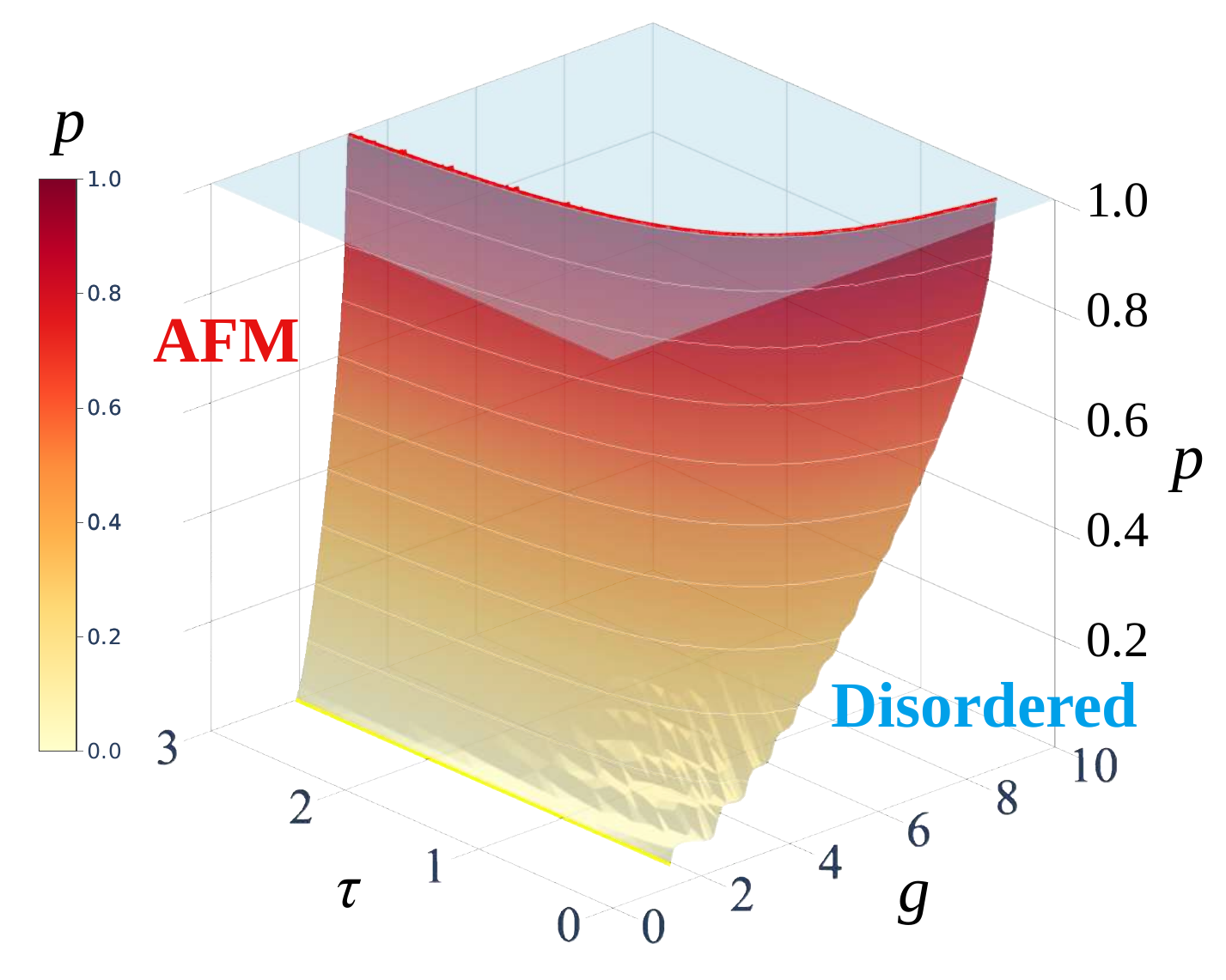}
\put (5,5) {{\textbf{(h)}}}
\end{overpic}
\caption{
For $(\tau,g)=(1, 3.5)$ in the MITE of the 2D CDHM:
(a) Binder ratios $R_{2,z}$ for different system sizes $L$ cross at $p_c \approx 0.354$.
(b) The inset panel shows $\langle |m_z| \rangle$ as a function of $p$ for various $L$, while the main panel presents the finite-size scaling and data collapse of $\langle |m_z|\rangle$. 
\\
For $(g,p)=(3, 0.1)$:
(c) Binder ratios $R_{2,z}$ for different system sizes $L$ cross at $\tau_c \approx 0.468$;
(d) $\langle |m| \rangle$ as a function of $\tau$ for various $L$ and data collapse of $\langle |m|\rangle$.
\\
For $(\tau,p)=(2,0.3)$:
(e) Binder ratios $R_{2,z}$ for different system sizes $L$ cross at $g_c \approx 2.8$; 
(f) $\langle |m| \rangle$ as a function of $g$ for various $L$ and data collapse of $\langle |m|\rangle$.
\\
(g) and (h): Critical surfaces viewed from different angles. 
}
\label{fig:cdhm_examples}
\end{figure*}

The parameter space of the steady state in this model is spanned by $(\tau, g, p)$.
Compared with the TFIM Hamiltonian, the CDHM Hamiltonian not only provides a 2D example but also exhibits a global continuous $SU(2)$ symmetry.
Interestingly, even at $g=0$, the ITE operator contains off-diagonal components $(S^+_iS^-_j+S^-_iS^+_j)/2$.
Increasing $g$ introduces both diagonal and off-diagonal terms, in sharp contrast to the TFIM case where tuning the Hamiltonian parameter $h$ only modifies the off-diagonal part. 
Following the same strategy as in the TFIM case, we fix two of the parameters in $\{\tau,g,p\}$ and vary the third to search for mixed-state phase transitions.

Taking the example of tuning $p$ at fixed $(\tau,g)=(1,3.5)$ in Fig.~\ref{fig:cdhm_examples}(a) and (b), we observe a clear crossing of the Binder ratio $R_{2,z}$ at $p_c \approx 0.354$, with $R_{2,z} \to 3$ for $p < p_c$ and $R_{2,z} \to 1$ for $p > p_c$. If $p=0$, the system is in the dimerized phase which has no order in the staggered magnetization. 
This indicates a mixed-state phase transition from a disordered phase to an AFM Ising-like ordered phase along the $z$-axis, driven by increasing the measurement strength.

Although the Binder ratio exhibits Ising-like characteristic, similar to the case of TFIM, the extracted critical exponents in the case of 2D CDHM again do not correspond clearly to any known universality class including the 3D Ising universality class for a closed system, as shown in Table~\ref{table:cdhm_exponents}. 
Similar phenomena can also be observed when tuning $g$ and $\tau$, as shown in Fig.~\ref{fig:cdhm_examples}(c-f) and Table~\ref{table:cdhm_exponents}.
Importantly, the critical exponents $\nu$ and $\beta$ also change as we tune $p$, $\tau$, or $g$. 

\begin{table}[t]
\centering
\begin{tabular}{c c c c c}
\toprule
Parameters & $L_{\min}$ & Critical point & $\nu$ & $\beta$ \\
\midrule

$(\tau,g)=(3.5,1)$
& 24 & $p_c=0.3537(4)$ & 0.714(7) & 0.650(11) \\
& 28 & $p_c=0.3537(4)$ & 0.716(6) & 0.651(11) \\
& 32 & $p_c=0.3538(5)$ & 0.717(6) & 0.650(12) \\

\midrule

$(g,p)=(3,0.1)$
& 24 & $\tau_c=0.4679(5)$ & 0.799(6) & 0.718(9) \\
& 28 & $\tau_c=0.4679(6)$ & 0.801(6) & 0.721(10) \\
& 32 & $\tau_c=0.4678(7)$ & 0.801(6) & 0.720(12) \\

\midrule

$(\tau,p)=(2,0.3)$
& 24 & $g_c=2.806(3)$ & 0.736(18) & 0.654(26) \\
& 28 & $g_c=2.806(2)$ & 0.739(17) & 0.653(26) \\
& 32 & $g_c=2.806(2)$ & 0.741(15) & 0.650(26) \\

\bottomrule
\end{tabular}

\caption{
Finite-size stability analysis of the critical exponents obtained from
data collapse of $\langle |m| \rangle$ in the CDHM.
System sizes are $L=24,28,32,40,48$.
Small system sizes are progressively discarded to test the stability
of the extracted critical exponents.
}

\label{table:cdhm_exponents}
\end{table}

We further extract dynamical exponents $z\approx 1$ at all three critical points; see Appendix~\ref{appx:cdhm_z}. The corresponding four-point cross-ratio analysis is presented in Appendix~\ref{appx:cdhm-cross-ratio}. Unlike the $(1+1)$D cylinder geometry, a spatial torus does not admit a simple analogue of the chord distance~\eqref{eq:chord-distance} or a reduction of the equal-time four-point function to a single conformal cross ratio. We therefore restrict the operator insertions to a $1$D periodic loop on the torus and employ the chord distance~\eqref{eq:chord-distance} as an approximate finite-size distance. With this prescription, the conventional $(2+1)$D TFIM at its quantum critical point exhibits a reasonably good data collapse consistent with the $3$D Ising CFT, whereas the MDITE critical point shows a noticeably poorer collapse. These results provide additional, although not conclusive, evidence that the finite-parameter MDITE critical point may not be described by a conventional CFT.

\subsection{Analysis of the phase transitions}
To further interpret these results, note that any nonzero measurement rate $p$ explicitly breaks the (weak) $SU(2)$ symmetry to a residual (weak) $U(1)\rtimes \mathbb{Z}_2$ symmetry, favoring the $z$-axis. 

Although increasing $g$ in the Hamiltonian introduces both diagonal and off-diagonal terms, only the diagonal components are compatible with the projective measurements along the $z$-axis, 
which correspond to the AFM Ising interactions. 
In Appendix.~\ref{appx:nooff}, we also present the results demonstrating the absence of off-diagonal long-range order in this model. 

To understand this kind of phase transitions, one can similarly construct an effective two-copy Hamiltonian in the continuous-parameter limit $\tau \to 0$, $p \to 0$, with $\tau \sim p$, as that in Sec.~\ref{sec:theoretical}.
In this limit, the resulting effective model corresponds to a bilayer geometry consisting of two CDHM layers coupled by interlayer Ising interactions. 
When each individual layer is in the dimerized disordered phase ($g>g_{c,\text{GS}}$), tuning the measurement rate $p$ (equivalently, the interlayer coupling strength) can drive Ising phase transitions associated with the 3D classical Ising universality class.

Away from the continuous limits, when $g> g_{c,\mathrm{GS}}$, and the imaginary-time length is short ($\tau g \lesssim p$), a high measurement rate $p$ suppresses the off-diagonal coherence generated by the ITE operator, leading to dynamics dominated by the diagonal AFM Ising terms. This drives the system into an ordered phase with nonzero staggered magnetization along the $z$-axis. 
In contrast, if $\tau g \gtrsim p$, the off-diagonal coherence builds up rapidly, and the final steady state remains disordered even at $p=1$. 

Therefore, for projective measurements along the $z$-axis, the MDITE with the 2D CDHM  generalizes the physics of the 1D TFIM discussed in Secs.~\ref{sec:result:tfim} and \ref{sec:theoretical} to 2D. 
The associated phase transitions also appear to lie beyond conventional universality classes.

Since the CDHM can also be simulated using cluster updates within QMC, the emergence of long-range orders induced by measurements can also be understood through the cluster formation in Appendix.~\ref{appx:qmc3}. We emphasize that it would be interesting to explore other types of decoherence channels that preserve other symmetries, such as the $SU(2)$ symmetry, which we leave for future work. In such cases, more advanced update algorithms beyond conventional cluster methods for QMC simulations would be necessary, potentially revealing qualitatively different types of phase transitions.

The schematic critical surface associated with this model is shown in Fig.~\ref{fig:cdhm_examples}(g) and (h). This critical surface also shows that large $g$ requires small evolution time, while long evolution time requires weak coupling $g$.  This inverse relationship between $g$ and $\tau$ closely resembles that of the 1D TFIM.

Moreover, we note that since the state still has a $U(1)$ symmetry, off-diagonal long-range order may emerge if the imaginary-time step length in the MDITE protocol is allowed to scale with the system size, i.e., $\tau \propto L^{\alpha}$ for a sufficiently large $\alpha$.
In this regime and for $g<g_{c,\text{GS}}$, the final step of the MDITE protocol (measurement followed by ITE) effectively reduces to a conventional ITE procedure for preparing the ground state or low-energy excited state of the Hamiltonian, with the initial state to be $\mathcal{E}_p(\rho_{n_d-1})$.
This regime represents an interesting extension of the present framework and may be explored in future studies.

\section{Conclusion and discussions}\label{sec:discussion}
In this work, we introduce MDITE as a channel-based setting for studying how imaginary-time filtering is
modified by decoherence. The protocol alternates ITE with outcome-averaged projective measurement channels, thereby generating a competition between low-energy filtering and local dephasing. 

As representative examples, we apply MDITE to the $1$D TFIM and the $2$D CDHM. In both cases, the resulting steady-state density matrix exhibits a mixed-state phase transition accompanied by numerical signatures of spontaneous symmetry breaking. These transitions can be diagnosed using conventional linear observables, including the magnetization and Binder ratio, making the phase structure directly accessible within the density-matrix formalism. Moreover, the four-point cross-ratio data do not exhibit a satisfactory collapse for either model, which suggests possible deviations from emergent conformal invariance at the finite-parameter MDITE critical points.

To clarify the physical origin of these transitions, we mapped the MDITE mixed
state to a doubled Hilbert space using the Choi--Jamiołkowski isomorphism. In
the continuous-parameter limit, this mapping gives a theoretically controlled effective
equilibrium description that captures the competition between imaginary-time
filtering and measurement-induced decoherence and qualitatively reproduces the
observed phase structure. The corresponding transitions fall into conventional
Ising universality classes. In the finite-parameter regime, however, the
extracted critical exponents differ substantially from the continuous-limit
Ising values and appear to depend on the protocol parameters. This suggests
critical behavior beyond the equilibrium benchmark, while a
complete analytical classification remains open.

Technically, we develop a diagrammatic representation of the evolving MDITE mixed
state, which provides a practical QMC framework for evaluating mixed-state
observables in interacting many-body systems. This formulation enables the
simulations reported here, including the 2D case, and can in
principle be applied to higher spatial dimensions when the underlying
Hamiltonian and channel remain sign-problem free. It can also be combined with
existing QMC techniques for computing information-theoretic quantities, such as
entanglement entropy
~\cite{hasting2010renyi,stephan2012renyi,melko2010renyi,jonathan2020renyi,Zhao2022renyi,ding2024reweight,wangz2025reweight,tarabunga2025bell,yan2023unlocking,li2024relevant,mao2025sampling,wang2025extracting}
and R\'enyi negativity or related partial-transpose moments
~\cite{ding2025tracking,Wukh2020negativity,Alba2013negativity,Wangfh2025negativity,chungchiamin2014negativity,yu2021pt,elben2020pt,Neven2021pt,tarabunga2025pt,liu2026entanglement}.

The present results motivate applying MDITE beyond the
projective-measurement channels studied here. Physically, the central question
is how different forms of many-body order are reshaped when imaginary-time
filtering is opposed by local decoherence. This question is also timely in view
of recent progress in quantum imaginary-time-evolution algorithms and
programmable quantum simulators, where non-unitary protocols, measurements, and
decoherence naturally coexist~\cite{Motta2020qite,McArdle2019vqite,Yuan2019theoryofvariational,Benedetti2021qite,Kondappan2023qite,Mao2023qite,Ding2024qite,Nishi2021qite-nisq,Cao2022qite,ZhangSX2024qite,Semeghini2021Probing,Lin2021QITE}. 
Applying MDITE to broader classes of
Hamiltonians, measurement bases, and decoherence channels could therefore
reveal mixed-state phases beyond the simple Ising-type ordering considered in
this work. Natural directions include mixed-state long-range entanglement and
topological order
~\cite{lu2023charac,lu2023longrange,Fanruihua2024diagnostics,Lu2024disentangling,sang2025markov,lessa2025higherformanomaly},
as well as strong-to-weak spontaneous symmetry breaking
~\cite{Lee2023decoh_criticality,Lessa2025swssb,sala2024swssb,gu2024swssb,huang2025swssb,zack2025renyi1corr,guo2025swssb,guoyc2025swssb,guoyc2025ldpo,zhang2025swssb,zpf2025swssb,Liu2025swssb,zhang2025swssbre,Sang2024approx,ding2026swssb}.
These directions would test how broadly the competition between
imaginary-time filtering and decoherence can organize mixed-state many-body
physics.

On the experimental side, MDITE could be implemented in platforms such as Rydberg atom arrays~\cite{Wu2021rydbergreview,Evered2023rydberg}, superconducting qubits~\cite{Wendin2017superconduct}, or trapped-ion systems~\cite{haffner2008trapped}. Such realizations would enable direct observation of the mixed-state criticalities predicted here, bridging theory and experiment.
Therefore, the MDITE framework may provide a useful platform for
designing quantum algorithms for
preparing nontrivial mixed states, potentially enabling efficient
simulations of open quantum many-body systems.

\section*{Acknowledgements}
We thank G.-Y. Zhu, B. Miao, Y. Guo, Z. Bi, C. Wang, and Y. Zou for helpful discussions. Zenan Liu thanks the China Postdoctoral Science Foundation under Grants No.2024M762935 and NSFC Special Fund for Theoretical Physics under Grants No.12447119. 
Zhe Wang thanks the China Postdoctoral Science Foundation under Grants No.2024M752898. 
This project is supported by the Scientific Research Project (No.WU2025B011) and the Start-up Funding of Westlake University.
The authors thank the high-performance computing center of Westlake University for providing HPC resources.

 \textbf{Data availability.}
Data are available on Zenodo~\cite{zenodo}.


\appendix

\section{A brief review of \\the stochastic-series-expansion QMC}\label{appx:qmc1}

The stochastic series expansion (SSE) method is a versatile and efficient quantum Monte Carlo approach for investigating both finite-temperature and ground-state properties of spin and bosonic systems.
It reformulates the partition function $Z=\Tr(e^{-\beta H})$ by performing a Taylor expansion of the exponential operator. 
The expansion is carried out in a chosen basis $\{|\alpha\rangle\}$, enabling efficient stochastic sampling of physical observables. 

By writing the Hamiltonian as a sum of local operators, 
\begin{equation}
H=-\sum_{a,b}H_{a,b},
\label{apeq2}
\end{equation}
where $a$ refers to the type of operators (diagonal or off-diagonal operator; site or bond operator) and $b$ marks the spatial indices (e.g., the labels of sites or bonds), we have  
\begin{align}
Z=&\sum_{\alpha}\langle\alpha|e^{-\beta H}|\alpha\rangle \nonumber \\
=&\sum_{\alpha}\sum_{\{H_{a,b}\}}\frac{\beta^n(M-n)!}{M!}\langle\alpha|\prod^{M}_{i=1}H_{a(i),b(i)}|\alpha\rangle,
\label{apeq1}
\end{align}
where $M$ is the truncation of the expansion order. 
To ensure that all operator sequences $\prod_{i=1}^M H_{a(i),b(i)}$ have the same fixed length $M$, we have introduced a null operator $H_{-1,-1}$ serves as an identity element filling the unoccupied positions in each sequence.


In the SSE method, each Monte Carlo step consists of a diagonal update and an off-diagonal update. The diagonal update alternates between diagonal operators and null operators while preserving the detailed balance condition.
Through this process, the total number of non-null operators $n$ in the operator sequence can change.
\begin{enumerate}[(a)]
    \item For each null operator, we try inserting a new diagonal operator at a random position with probability
    \begin{equation}
        P_{\mathrm{add}}(n\rightarrow n+1)=\min\bigg\{\frac{\beta N_b\langle \alpha| H_b|\alpha\rangle}{M-n},1\bigg\},
    \label{apeq3}
    \end{equation}
    where $N_b$ denotes the total number of spatial bonds (or sites) available for insertion.
    \item For each diagonal operator, we remove it with probability
    \begin{equation}
        P_{\mathrm{remove}}(n+1\rightarrow n)=\min\bigg\{\frac{M-n+1}{\beta N_b\langle \alpha| H_b|\alpha\rangle},1\bigg\}.
    \label{apeq4}
    \end{equation}
\end{enumerate}


The off-diagonal update will change the number of operators and update between diagonal operators and off-diagonal operators. 
To improve the update efficiency, some nonlocal update scheme such as the loop update, cluster update, or directed loop update are typically used~\cite{Sandvik2010lecture,sandvik2003stochastic,Anders2002Quantum}. 

Next, we take the TFIM in Sec.~\ref{sec:result:tfim} as an example to illustrate the SSE method, and it is similar to discuss the CDHM in Sec.~\ref{sec:result:cdhm}. The Hamiltonian of the TFIM is given by 
\begin{equation}
H=  -J\sum_{\langle ij\rangle}Z_i Z_j -h\sum_{i} X_i .
\end{equation}
To express this Hamiltonian in the form of Eq.~\eqref{apeq2}, we introduce the following operators:
\begin{align}\label{eq:appx:tfim_op}
H_{0,i}&\equiv h I_i ,\nonumber\\
H_{1,i}&\equiv hX_i ,\nonumber \\
H_{2,\langle ij\rangle}&\equiv J(Z_iZ_j+I_iI_j ).
\end{align}
In the computational basis, defined as the eigenbasis of the Pauli-$Z$ operators, $H_{0,i}$ represents a diagonal on-site operator associated with site $i$, $H_{1,i}$ is an off-diagonal on-site operator, and $H_{2,\langle ij\rangle}$ is a diagonal bond operator associated with the site pair $\langle ij\rangle$.
The nonzero matrix elements of the operators defined in Eq.~\eqref{eq:appx:tfim_op} are 
\begin{align}
\langle\uparrow_i| H_{0,i} |\uparrow_i\rangle&=\langle\downarrow_i| H_{0,i} |\downarrow_i\rangle=h \nonumber\\
\langle\uparrow_i| H_{1,i} |\downarrow_i\rangle&=\langle\downarrow_i| H_{1,i} |\uparrow_i\rangle=h \nonumber\\
\langle\uparrow_i\uparrow_j| H_{2,\langle ij\rangle} |\uparrow_i\uparrow_j\rangle&=\langle\downarrow_i\downarrow_j | H_{2,\langle ij\rangle} |\downarrow_i\downarrow_j\rangle=2J
\end{align}

\begin{figure}[ht!]
\centering
\begin{overpic}[width=0.4\textwidth]{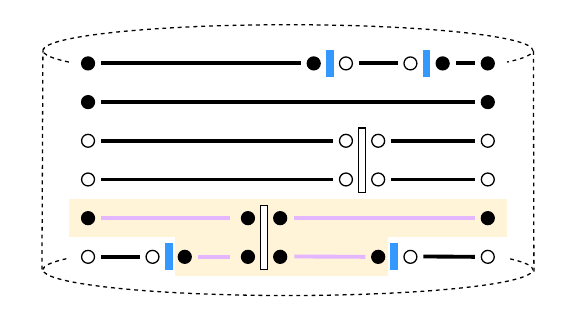}
\end{overpic}
\caption{The diagram for cluster update in the TFIM model.  White rectangle and blue rectangle denotes the diagonal operator and site operator respectively. The yellow area represents a constructed cluster, with the purple solid line indicating the update line. The line stops upon encountering a site operator; when it meets a bond operator, it continues to extend branches from the other legs of that operator. The dashed lines represent periodic boundary conditions in imaginary time.
}
\label{fig:cluster}
\end{figure}

For the cluster update, we construct clusters according to the following two rules: (i) A cluster is terminated by a site operator; (ii) The update line continues to extend from the other legs of a bond operator, while each bond operator belongs to exactly one cluster. The process is repeated until all clusters are identified. An illustration of the cluster construction is shown in Fig.~\ref{fig:cluster}, where the yellow area is a cluster. Finally, each cluster is flipped with probability $1/2$. As we will show below, this update can be readily adapted to the simulation of the extended ensemble in Eq.~\eqref{eq:pt_target}.

\section{Diagrammatic representation of the MDITE mixed states}\label{appx:pathint}
Here, we introduce a diagrammatic representation to describe the evolving state $\rho_{n_d}$ in the MDITE process. 
This representation is powerful as it both enhances conceptual understanding and facilitates numerical studies, such as the efficient QMC method we introduce for the MDITE problem in Sec.~\ref{sec:qmc}. 

For convenience, we assume an orthonormal and local computational basis ${\ket{s}}$, with $s\in\{0,1\}^{\otimes N}$, consistent with the framework of quantum circuits and standard QMC simulations.

\subsection{ITE without measurements}\label{sec:ite_no_measure}
\begin{figure}[ht]
    \centering
    \subfigure[]{
        \includegraphics[height=0.134\textwidth]{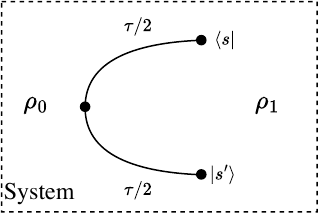}}
    \subfigure[]{
        \includegraphics[height=0.134\textwidth]{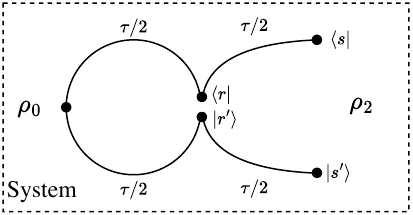}}
    \caption{
        Diagrammatic representation of the MDITE at measurement rate $p=0$, with the input (left) state given by $\rho_0 = I_{2^N} / 2^{N}$. 
        Each node (black filled circle) represents a basis state. 
        The label ``System'' indicates that the diagram applies to all qubits in the entire system.
        (a) For the first time step, after an evolution time of $\tau/2$, the initial state $\rho_0$ evolves into $\rho_1 \propto e^{-\tau H}$ on the right, where $\bra{s}$ and $\ket{s'}$ denote the bra and ket components of $\rho_1$ in Eq.~\eqref{eq:rho1}, respectively. 
        (b) Starting from $\rho_1$ with bra and ket components $\bra{r}$ and $\ket{r}$, another evolution for time $\tau/2$ (i.e., in the second time step) yields the output state $\rho_2 \propto e^{-2\tau H}$, where $\bra{s}$ and $\ket{s'}$ correspond to the components of $\rho_2$.
    }
    \label{fig:diagram_no_measure}
\end{figure}

We begin by examining the case where no measurements are applied throughout the ITE processes, namely when the measurement rate is set to $p=0$. 
The state after the first time step is
\begin{equation}
    \rho_1 \propto e^{-\frac{\tau}{2}H}I_{2^N}e^{-\frac{\tau}{2} H}=e^{-\tau H},
\end{equation}
which can be written as 
\begin{equation}\label{eq:rho1}
    \rho_1 = \sum_{s,s'} \langle s'|\rho_1|s\rangle \ket{s'}\bra{s}
\end{equation}
in the computational basis $\{\ket{s}\}$. 
Fig.~\ref{fig:diagram_no_measure}(a) presents a diagrammatic representation of $\rho_1$, 
showing that an initial state $\rho_0=I_{2^N}/2^N$ evolves into $\rho_1$, which is spanned by $\ket{s'}\langle s|$, under imaginary-time propagation of duration $\tau/2$.  
For clarity, the summation over different $\ket{s'}\bra{s}$ in Eq.~\eqref{eq:rho1} is omitted in the diagram.

Similarly, the output state after the second step, $\rho_2 \propto e^{-2\tau H}$, is depicted in Fig.~\ref{fig:diagram_no_measure}(b).
We note that the intermediate states $\ket{r'}\bra{r}$ in the representation of $\rho_2$ [Fig.~\ref{fig:diagram_no_measure}(b)] correspond to $\ket{s'}\bra{s}$ in the representation of $\rho_1$ [Fig.~\ref{fig:diagram_no_measure}(a)], respectively.  
This follows from $\rho_2 \propto e^{-\tau/2} \rho_1 e^{-\tau/2}$, where the two ITE operators act on the upper and lower sides of $\rho_1$ in  Fig.~\ref{fig:diagram_no_measure}(a).
In a manner analogous to tensor-network contraction, these operators propagate $\rho_1$, represented by $\ket{r'}\bra{r}$, into $\rho_2$, represented by $\ket{s'}\bra{s}$. 

Notably, Fig.~\ref{fig:diagram_no_measure}(a) and (b) can be interpreted as two ensembles, due to the summation over all possible $\ket{s'}\bra{s}$ in Eq.~\eqref{eq:rho1}. 
The cases of $\rho_k$ with $k>2$ follow in complete analogy, with their diagrammatic representations obtained recursively in the same way.

\subsection{ITE with deterministic measurements}\label{sec:measure_all}
We next consider the case in which all qubits are measured deterministically in each time step.  
Note that the output state $\rho_1 \propto e^{-\tau H}$ after the first time step is independent of the choice of measurement channel $\mathcal{E}_p$ and whether $\mathcal{E}_p$ is applied, since measurements on a maximally mixed state always leave it unchanged. Therefore, the diagrammatic representation of $\rho_1$ under deterministic measurements remains Fig.~\ref{fig:diagram_no_measure}(a).

\begin{figure}[ht]
    \centering
    \subfigure[]{
        \includegraphics[height=0.134\textwidth]{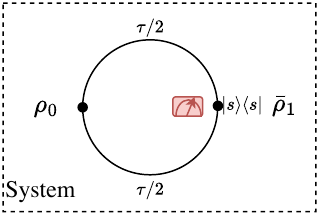}}
    \subfigure[]{
        \includegraphics[height=0.134\textwidth]{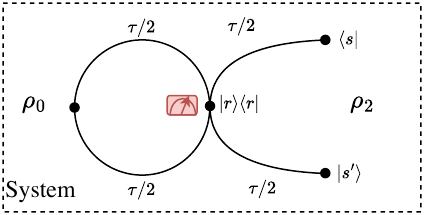}}
    \caption{
        Diagrammatic representations of the MDITE under deterministic projective measurements. 
        (a) Compared with Fig.~\ref{fig:diagram_no_measure}, the measurement on $\rho_1$ identifies the bra and ket components of $\rho_1$ in Eq.~\eqref{eq:rho1}, forming a closed loop for the measurement-averaged state $\bar{\rho}_1$.
        (b) The output state $\rho_2$ after the second time step.
    }
    \label{fig:diagram_measure}
\end{figure}

In the second time step, we first obtain the measurement-averaged state after the measurement channel $\mathcal{E}_p$, which is 
\begin{equation}\label{eq:rho1_determ}
    \bar{\rho}_1 = \sum_{s} \langle s|\rho_1|s\rangle \ket{s}\bra{s}.    
\end{equation}
Compared with $\rho_1$ in Eq.~\eqref{eq:rho1}, the projective measurements identify the bra state $\bra{s}$ and the ket state $\ket{s'}$; that is, the $\bar\rho_1$ is obtained from $\rho_1$ by keeping only the diagonal matrix elements. As a result, the diagrammatic representation of $\bar{\rho}_1$ forms a closed loop, as illustrated in Fig.~\ref{fig:diagram_measure}(a). 

After the ITE, the resulting state is
\begin{equation}
    \rho_2 \propto e^{-\tfrac{\tau}{2}H}\,\bar{\rho}_1\,e^{-\tfrac{\tau}{2}H},
\end{equation} 
which is the output state after the second time step. Analogous to the discussion in Sec.~\ref{sec:ite_no_measure}, $\rho_2$ can be represented diagrammatically as in Fig.~\ref{fig:diagram_measure}(b). 
Here, the state $\ket{r}\bra{r}$ corresponds to the state $\ket{s}\bra{s}$ in $\bar{\rho}_1$ shown in Fig.~\ref{fig:diagram_measure}(a).
It is straightforward to generalize the discussion to a general $\rho_k$ $(k>2)$, and the diagrammatic representations also follow recursively.

\subsection{ITE with deterministic measurements on subsystem $A$ only}\label{sec:a_measure}
As the final preparatory step before discussing the general case of probabilistic measurements on each qubit, we consider the scenario in which all qubits in subsystem $A$ are measured projectively with probability one, while the complementary subsystem $B \equiv \overline{A}$ remains unmeasured.

For convenience, we rewrite the local basis states $\{\ket{s}\}$ as $\{\ket{s_A}\otimes\ket{s_B}\}$ or $\{\ket{s_A,s_B}\}$.
Then the measurement-averaged state of $\rho_1$ in the second time step can be written as 
\begin{equation}\label{eq:deduct1}
        \bar{\rho}_1
        =\sum_{s_A,s_B,s_B'}\langle s_A,s_B'|\rho_1|s_A,s_B\rangle 
         \ket{s_A,s_B'}\bra{s_A,s_B},
\end{equation}
where $s_A \in \{0,1\}^{\otimes N_A}$ and $s_B \in \{0,1\}^{\otimes N_B}$, with $N_A$ and $N_B$ denoting the number of qubits in subsystems $A$ and $B$, respectively.

\begin{figure}[ht]
    \centering
    \subfigure[]{
        \includegraphics[height=0.134\textwidth]{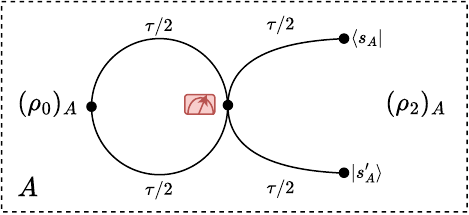}}
        \subfigure[]{
        \includegraphics[height=0.134\textwidth]{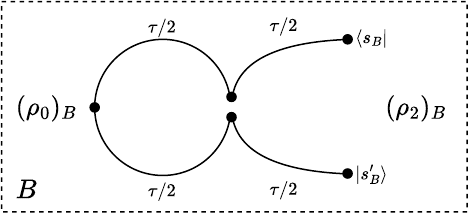}}
    \caption{
        Diagrammatic representation of the MDITE where only subsystem $A$ is projectively measured in a deterministic manner. 
        The label ``$A$'' (``$B$'') indicates that the diagram applies to all qubits in subsystem $A$ ($B$).
        Since the computational basis is local, the evolutions of the reduced density matrices $(\rho_0)_A = \Tr_B(\rho_0)$ and $(\rho_0)_B = \Tr_A(\rho_0)$ can be depicted separately in (a) and (b), respectively.
    }
    \label{fig:diagram_measure_A}
\end{figure}

Since the computational basis is local, the basis state $\ket{s_A}$ of subsystem $A$, which is measured, in Eq.~\eqref{eq:deduct1} behaves exactly as $\ket{s}$ in Eq.~\eqref{eq:rho1_determ}. Similarly, for subsystem $B$, the states $\ket{s_B}$ and $\ket{s_B'}$ behave exactly as $\ket{s}$ and $\ket{s'}$ do in Eq.~\eqref{eq:rho1}, where the measurement rate is zero for all qubits.
Consequently, the diagrammatic representation of $\rho_2 \propto e^{-\frac{\tau}{2}H} \, \bar{\rho}_1 \, e^{-\frac{\tau}{2}H}$ in this case can be represented by Fig.~\ref{fig:diagram_measure_A}, with two components corresponding to subsystems $A$ and $B$. 
Likewise, the diagrammatic representation for $\rho_k$ with $k>2$ is omitted here, since the generalization is natural.

\subsection{ITE with probabilistic measurements}\label{sec:measure_prob}
Now we extend the discussions above to a general case in which each site is independently measured with probability $p$ in each time step.

For a system of $N$ qubits and $k$ time steps, the binary choice of measuring or not measuring each qubit in each layer yields $2^{(k-1)N}$ distinct ensembles, which together form an \emph{extended ensemble}. 
In this sense, the examples in Figs.~\ref{fig:diagram_no_measure}(b), \ref{fig:diagram_measure}(b), and \ref{fig:diagram_measure_A} correspond to three of the $2^N$ possible ensembles when $k=2$. 

\begin{figure*}[ht]
    \centering
    \subfigure[]{
        \includegraphics[width=0.45\textwidth]{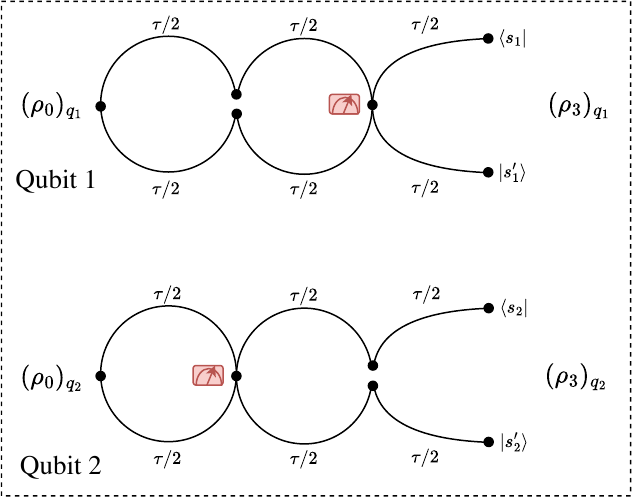}}
    \subfigure[]{
        \includegraphics[width=0.45\textwidth]{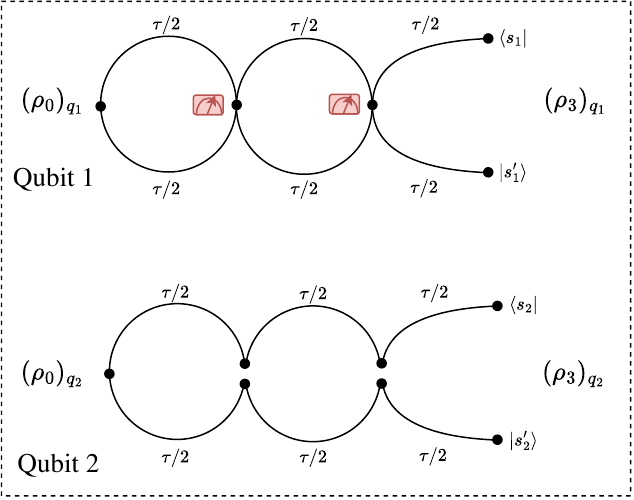}}
    \caption{
        For $N=2$, $k=3$, and $\{\ket{s} \equiv \ket{s_1, s_2}\}$, diagrammatic representations of two possible ensembles are shown. 
        The subscripts $q_1$ and $q_2$ denote the reduced density matrices of qubit 1 and qubit 2, respectively. 
        (a) In the second layer, qubit 1 is unmeasured while qubit 2 is measured; in the third layer, qubit 1 is measured while qubit 2 is unmeasured. 
        (b) Qubit 1 is measured in both the second and third layers, whereas qubit 2 remains unmeasured throughout.
    }
    \label{fig:diagram_prob}
\end{figure*}

The diagrammatic representation with probabilistic measurements now consists of $N$ components, a natural generalization of Fig.~\ref{fig:diagram_measure_A}, one for each qubit. 
In Fig.~\ref{fig:diagram_prob}, we present an additional example for $N=2$ and $k=3$, illustrating two of the $2^4$ possible ensembles.

\subsection{QMC simulations}\label{sec:qmc}
Importantly, the diagrammatic representation of the evolving state $\rho_k$ naturally facilitates QMC studies of $\rho_{n_d}$ based on imaginary-time path integrals (after applying $n_d$ layers, we obtain the final output state $\rho_{n_d}$), such as the stochastic series expansion (SSE)~\cite{Sandvik1992sse,Sandvik1999sse,Sandvik2010lecture,syljuaasen2002quantum,sandvik2019stochastic,Yan2019sweeping,yan2022global}, as long as the Hamiltonian is sign-problem-free. 
In addition, it also makes it straightforward to implement tensor-network–based simulations~\cite{Ran2020tensor}.

The QMC simulations can be carried out by sampling configurations according to the generalized partition function $\mathcal{Q}_{n_d}\propto \Tr(\rho_{n_d})$, which is associated with the extended ensemble.
For a specific ensemble, suppose that in the $k$th ($k \ge 2$) time step, only the qubits in subsystem $A_k$ are measured. 
We denote the corresponding generalized partition function for this ensemble by $Q_{n_d}(\{A_k\})$. For instance, when $n_d=2$, the generalized partition function for the ensemble depicted in Fig.~\ref{fig:diagram_measure_A} is written as $Q_{2}(A_2\equiv A)$, while further identifying $\ket{s}$ and $\ket{s'}$ components because of the trace operation in $\mathcal{Q}_{n_d}$.

Therefore, we have 
\begin{equation}\label{eq:pt_target}
    \mathcal{Q}_{n_d}=\sum_{\{A_k\}}\prod_{l=2}^{n_d}
    \bigg[
        p^{N_{A_l}}(1-p)^{N-N_{A_l}}  
    \bigg]Q_{n_d}(\{A_k\}),
\end{equation}
where $N_{A_l}$ denotes the number of qubits in subsystem $A_l$. Since different ensembles $\{Q_{n_d}(\{A_k\})\}$ carry different weights determined by the measurement rate $p$, transitions between ensembles can be implemented in QMC simulations by designing appropriate selection probabilities that satisfy the detailed balance condition.

Within the SSE framework, we have developed an unbiased and efficient QMC algorithm to simulate Eq.~\eqref{eq:pt_target}, applicable to Hamiltonians for ITEs in arbitrary spatial dimensions. While the present work focuses on the steady state $\rho_{\infty}$ ($n_d\to\infty$), the algorithm also enables studies of the dynamics of the MDITE process. Further details of the QMC algorithm are provided in Appendices~\ref{appx:qmc1} and \ref{appx:qmc2}.

\section{QMC algorithm for simulating the generalized partition function $\mathcal{Q}_{n_d}$}\label{appx:qmc2}
\begin{figure}[ht!]
\centering
\begin{overpic}[width=0.45\textwidth]{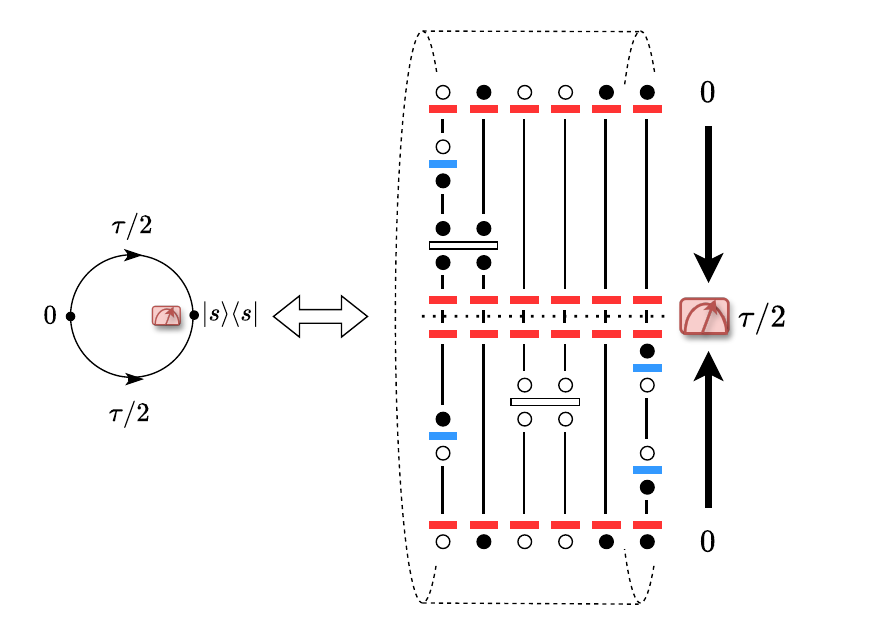}
\end{overpic}
\caption{ 
The measurement-averaged state $\overline{\rho}_1$ corresponds to the sampling configuration in the SSE stimulation. 
In the diagram, white rectangles denote diagonal operators, blue rectangles represent site operators, and red rectangles indicate auxiliary identity operators.
Additional auxiliary identity operators (red rectangles) are introduced at each local site. 
Those auxiliary operators do not affect the physical properties of the system, and they facilitate the treatment of boundary conditions in simulations practically. 
}
\label{fig:ssem}
\end{figure}

We still take the TFIM as an example. Compared with the standard partition function $Z$, the simulation of $\mathcal{Q}_{n_d}$ involves multiple replicas and probabilistic boundary conditions that connect them. Fig.~\ref{fig:ssem} shows the SSE configuration of $\bar{\rho}_1\propto e^{-\tau H}$, whose QMC simulation is equivalent to that of a Gibbs state with partition function $Z = \Tr(e^{-\tau H})$.

\begin{figure}[htbp]
    \centering
    \subfigure[]{\includegraphics[width=0.4\textwidth]{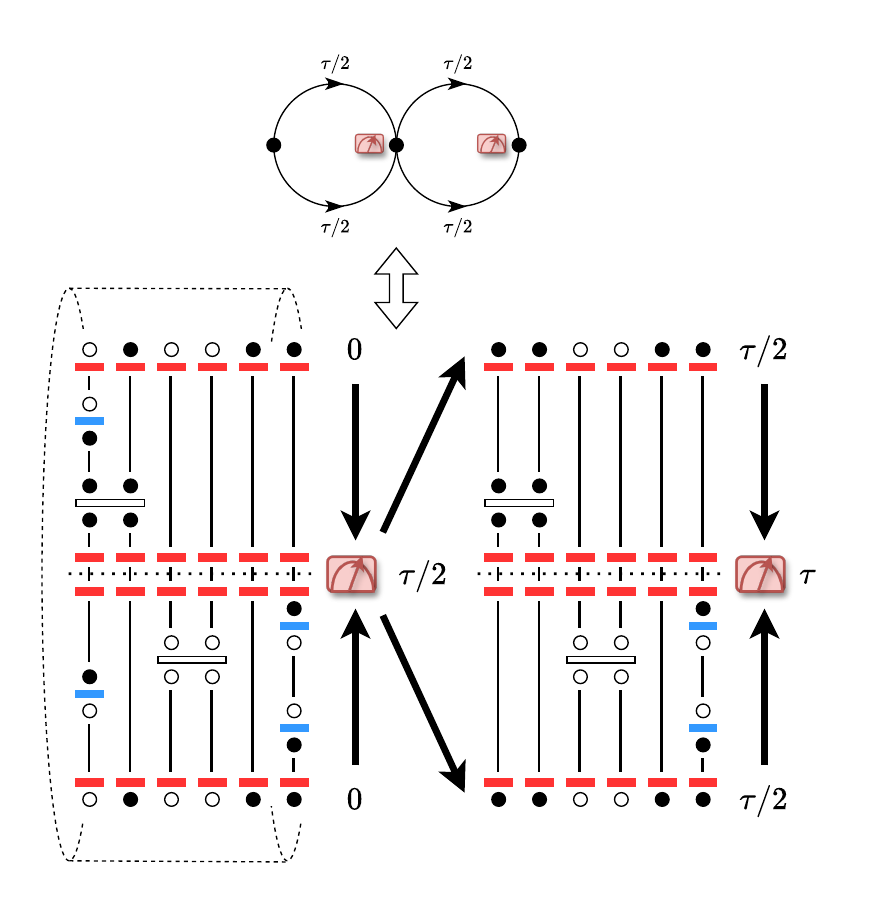}}
    \subfigure[]{\includegraphics[width=0.4\textwidth]{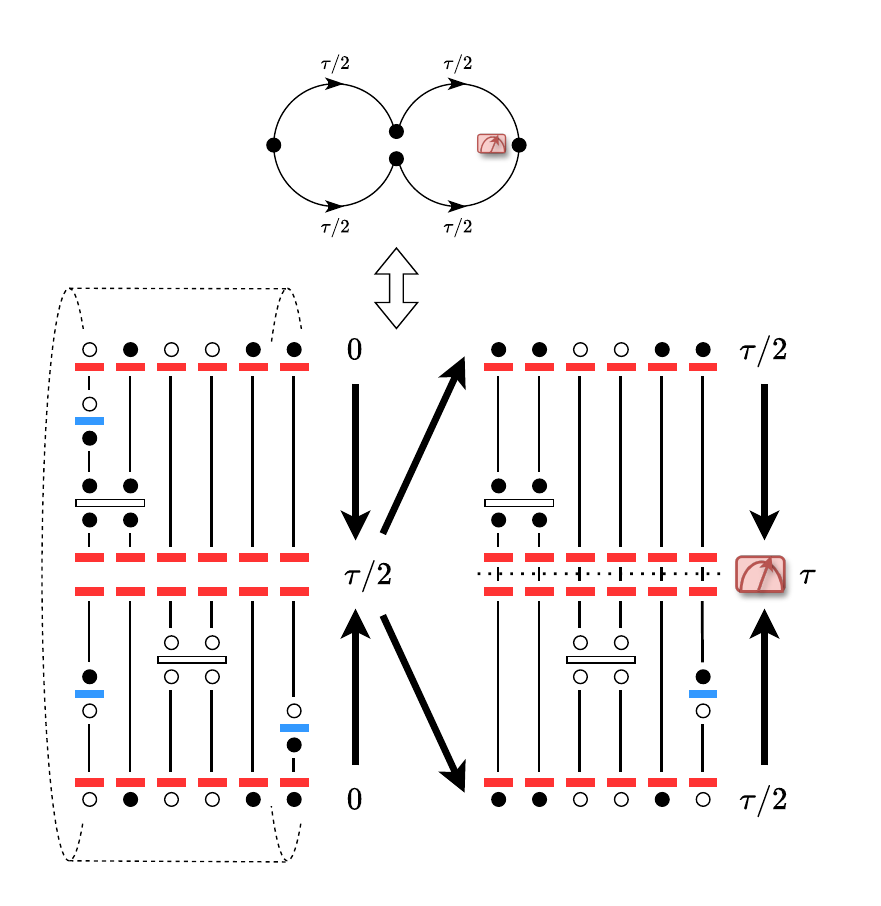}}
    \caption{
        The measurement-averaged state $\overline{\rho}_1$ corresponds to the sampling configuration in the SSE stimulation, taking two time steps as an example. White rectangle and blue rectangle represents the diagonal operator and site operator respectively, while red rectangle denotes the auxiliary identity operators. In (a), the simulation performs a measurement operation after a time interval of $\tau/2$. At this moment, the manifold closes, and the two time steps are adhered together. In (b), no measurement is performed at $\tau/2$, which the manifold remains open, and the two circuits form a single larger layer. The dashed lines represent the periodic boundary conditions in imaginary time. The two subfigures correspond respectively to the two extreme cases of measurement rates $p = 1$ and $p = 0$.
    }
    \label{fig:ssem2}
\end{figure}

Similarly, when the measurement rate is $p=1$, as discussed in Sec.~\ref{sec:measure_all}, the SSE configuration related to $\Tr(\rho_2)$ is illustrated in Fig.~\ref{fig:ssem2}(a). 
Since the state undergoes two evolutions, the diagram contains two replicas of $e^{-\tau H}$. 
To perform the QMC simulations, we note that the boundary conditions must be consistent with the diagrammatic representation introduced in Sec.~\ref{appx:pathint}, which imposes constraints on the spin configurations near the boundaries of each replica.
Since the diagonal update in SSE is local and does not alter the spin states, it can be implemented in the same manner as that in a standard partiton function $Z$. 
For the nonlocal update, the boundary conditions require that the spins at the interfaces between different replicas remain consistent in Fig.~\ref{fig:ssem2}(a).
Consequently, when a cluster extends across an interface, it may continue into another replica.
In practice, this is implemented using auxiliary identity operators: when two replicas are connected, their corresponding auxiliary identity operators are required to belong to the same cluster.

When the measurement rate is zero, the ensemble reduces to the standard partition function with a doubled inverse temperature, as illustrated in Fig.~\ref{fig:ssem2}(b). In this case, the auxiliary operators are not involved in the cluster update.

As discussed in Sec.~\ref{sec:measure_prob}, for a general measurement rate $p$, we only need to switch the simulated ensemble between Fig.~\ref{fig:ssem}(a) and (b) when the number of time steps is two.
Specifically, before each Monte Carlo step, we consider the following merge-split process:
\begin{enumerate}[(1)]
    \item  If two spins near the boundary in the corresponding replicas are in the same state but are not connected in the diagrammatic representation, they are merged with probability
        \begin{equation}
        P_{\mathrm{merge}}=\min\bigg\{\frac{p}{1-p},1 \bigg\};
        \end{equation}
    \item If two spins are in the same state and already connected in the diagrammatic representation, they are split with probability
\begin{equation}
P_{\mathrm{split}}=\min\bigg\{\frac{1-p}{p},1 \bigg\}.
\end{equation}
\end{enumerate}

The merge-split process can be naturally generalized to the case when the number of time steps is greater than two, i.e., more replicas of $e^{-\tau H}$, following the dicussions in Sec.~\ref{sec:qmc}.
For the Heisenberg model, the diagonal update, merge–split process, and nonlocal update are implemented in a manner analogous to those used for the TFIM under measurement. The cluster update scheme can be employed to perform updates between diagonal and off-diagonal operators in the Heisenberg model.

\section{Cluster formations and long-range orders}\label{appx:qmc3}
In this appendix, we provide an intuitive, QMC-based picture for how local projective measurements enhance collective behavior in the system, ultimately generating a mixed ordered phase.

As introduced in Appendix.~\ref{appx:qmc1}, the nonlocal cluster update is a crucial step to ensure the ergodicity of the QMC algorithm for the TFIM. Therefore, the size of the automatically formed clusters reflects the degree of collective behavior among spins: large clusters emerge when many spins fluctuate coherently.
For example, in the 1D TFIM, when the transverse field is very strong $(h\to\infty)$, spins are dominated by the external field, and cluster updates produce only small, local clusters. 
Conversely, when the field is very weak, $(h\to 0)$, the Ising interactions dominate, connecting all sites into a single large clusters. In this regime, cluster construction naturally produces system-spanning clusters, reflecting the collective alignment of spins in the FM phase.

According to the merge-split process introduced in Appendix.~\ref{appx:qmc2}, the presence of measurements allows update lines to connect across replicas, effectively merging clusters from different replicas. Therefore, starting with the PM ground state of the TFIM, even local measurements can induce the formation of larger clusters during the QMC updates, which is similar to increasing the Ising coupling strength for the ground state of the TFIM. This reflects why measurements can drive the system into a mixed-ordered phase with long-range order, from the perspective of cluster formations in QMC simulations. 
Similar reasoning also applies to the CDHM, where a cluster update is also used. 

We emphasize that for general Hamiltonians, the specific form of the nonlocal update scheme may differ from the cluster update employed in the TFIM and CDHM. Depending on the model, the corresponding update mechanism could also reveal new insights into the intrinsic properties of the states in the MDITE process.

\section{Fittings of the other two dynamical exponents for the 1D TFIM}\label{appx:tfim_z}
In this appendix we present additional numerical results for the extraction of the dynamical exponent $z$ in the example of 1D TFIM in Sec.~\ref{sec:dynamical_tfim}, using Eq.~\eqref{eq:fit_z}.
These results complement the analysis presented in Sec.~\ref{sec:dynamical_tfim} and further support the conclusion that the mixed-state critical points exhibit a dynamical exponent close to $z=1$. 
See Table~\ref{table:tfim_z}.

\begin{table}[htb]
\centering
\begin{tabular}{c c c c}
\toprule
$(\tau,h,p)$ & $L_{\min}$ & $a$ & $z$ \\
\midrule

(1.2,1.8373,0.8)
& 64 & 9.2(2) & 1.013(5) \\
& 72 & 9.1(3) & 1.010(6) \\
& 80 & 9.0(3) & 1.009(7) \\
& 88 & 8.6(4) & 0.997(9) \\
& 96 & 8.1(5) & 0.99(1) \\

\midrule

(0.2651,2.5,0.5)
& 64 & 2.2(1) & 0.97(1) \\
& 72 & 2.2(1) & 0.97(1) \\
& 80 & 2.0(1) & 0.96(1) \\
& 88 & 1.8(2) & 0.93(2) \\
& 96 & 2.0(2) & 0.96(3) \\

\bottomrule
\end{tabular}

\caption{
Extraction of the dynamical exponent $z$ from the relaxation time 
for two critical points in the example of TFIM.
System sizes range from $L=64,72,..., 128$.
To test the stability of the extracted exponent,
small system sizes are gradually discarded.
$L_{\min}$ denotes the smallest system size included in the fits.
}

\label{table:tfim_z}
\end{table}

\section{Dynamical scaling of the example of 2D CDHM}\label{appx:cdhm_z}
\begin{figure}[htbp]
    \centering
    \subfigure[]{\includegraphics[width=0.35\textwidth]{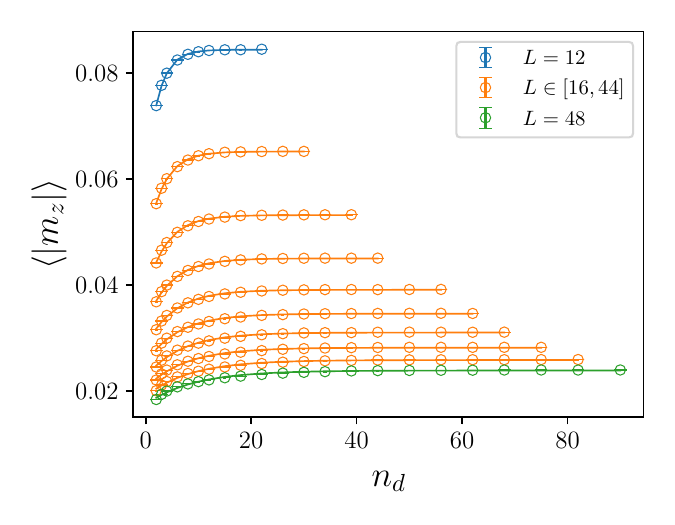}}
    \subfigure[]{\includegraphics[width=0.35\textwidth]{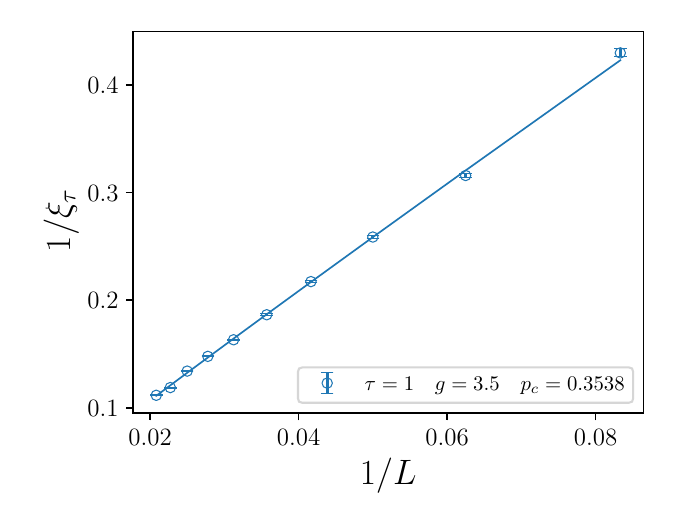}}
    \caption{
        When setting $(\tau,g,p)=(1,3.5,0.3538)$ for the MDITE with the 2D CDHM: (a)  The relaxation process of $\langle |m| \rangle$ with increasing the number of time steps $n_d$ for various system sizes $L$. It satisfies the formula $m=a\exp(-n_d/\xi_\tau)+b$, where $a$, $\xi_\tau$ and $b$ are fitting parameters.
(b) Extracting the dynamical exponent $z$ from $\xi_\tau \sim L^z$ with various system size $L$.
    }
    \label{fig:2d_cdhm_z}
\end{figure}
Similar to the example of 1D TFIM, we extract the dynamical exponent $z$ for the 2D CDHM by simulating the relaxation dynamics of the order parameter $\langle |m_z| \rangle$. 

As a representative case, for $(\tau,g)=(1,3.5)$ the phase transition occurs at $p_c \approx 0.3538$, as shown in Fig.~\ref{fig:cdhm_examples}(a) and (b). 
From the relaxation behavior illustrated in Fig.~\ref{fig:2d_cdhm_z}, we extract the relaxation time $\xi_\tau$ and obtain a dynamical exponent $z \approx 0.963$, which is close to unity. 
The stability analysis further confirms $z \approx 1$, as summarized in Table~\ref{table:cdhm_z}.

We also determine the dynamical exponent for two additional parameter sets, $(\tau,g,p)=(2,2.806,0.3)$ and $(0.4679,3.0,0.1)$. The extracted values again yield $z \approx 1$. 


\begin{table}[htb!]
\centering
\begin{tabular}{c c c c c}
\toprule
$(\tau,g,p)$ & $L_{\min}$ & $a$ & $z$ \\
\midrule

(1,3.5,0.3538) 
& 12 & 4.63(6) & 0.963(4) \\
& 16 & 4.56(7) & 0.959(4) \\
& 20 & 4.64(8) & 0.963(5) \\
& 24 & 4.6(1)  & 0.963(6) \\

\midrule

(2,2.806,0.3)
& 8  & 11.3(3) & 1.092(6) \\
& 12 & 11.3(3) & 1.093(6) \\
& 16 & 11.2(3) & 1.091(7) \\
& 20 & 11.4(3) & 1.096(8) \\

\midrule

(0.4679,3.0,0.1)
& 12 & 2.17(6) & 1.015(8) \\
& 16 & 2.17(6) & 1.015(8) \\
& 20 & 2.14(7) & 1.011(8) \\
& 24 & 2.08(8) & 1.00(1) \\

\bottomrule
\end{tabular}

\caption{
Extraction of the dynamical exponent $z$ from the relaxation time in the CDHM using system sizes $L\in [8,48]$.
Different parameter sets $(\tau,g,p)$ correspond to three quantum critical points.
To test the stability of the fits, small system sizes are gradually discarded.
The resulting values are consistently close to $z\approx 1$.
}

\label{table:cdhm_z}
\end{table}

\section{Four-point correlation function of cross ratios for 2D CDHM}\label{appx:cdhm-cross-ratio}
For a $(2+1)$D system with spatial torus geometry, there is no simple analogue of the chord distance on a circle, and the equal-time four-point function cannot generally be reduced to a function of a single real cross ratio. As an approximate finite-size test, we therefore place the four operator insertions along a $1$D periodic loop of the torus and evaluate the distances using Eq.~\eqref{eq:chord-distance}. We then construct the corresponding lattice four-point function and its normalized counterpart following the procedure introduced for the $1$D models.

\begin{figure}[htbp]
    \centering
    \subfigure[]{
        \includegraphics[width=0.30\textwidth]{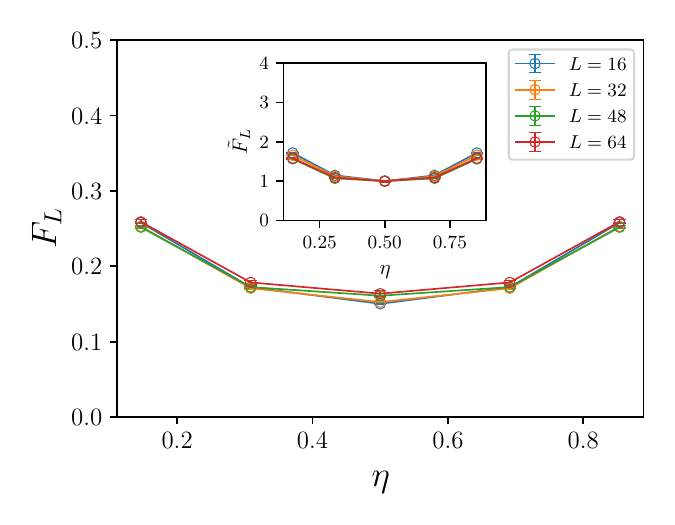}
    }
    \subfigure[]{
        \includegraphics[width=0.30\textwidth]{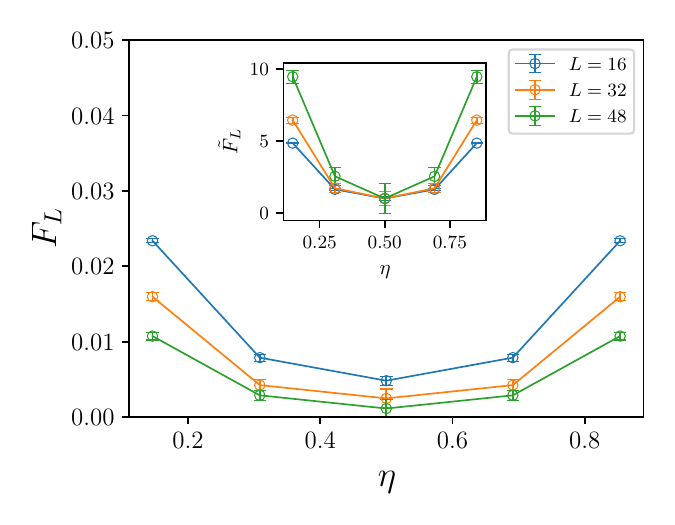}
    }
    \caption{
        Approximate four-point cross-ratio analysis for
         (a) the conventional $(2+1)$D TFIM at its quantum critical point and
         (b) the MDITE steady state of the $2$D CDHM at
        $(\tau,g,p)=(1,3.5,0.3538)$.
        The TFIM is included as a benchmark for the approximate construction and exhibits a reasonably good collapse, whereas the MDITE data show a poorer collapse with noticeable residual system-size dependence.
        The scaling dimensions used in the two panels are determined independently from the corresponding finite-size scaling analyses.
    }
    \label{fig:four-point-2d}
\end{figure}

Fig.~\ref{fig:four-point-2d} compares the approximate four-point cross-ratio functions for (a) the conventional $(2+1)$D TFIM at its quantum critical point and  (b) the MDITE steady state of the $2$D CDHM at $(\tau,g,p)=(1,3.5,0.3538)$. The TFIM serves as a benchmark for assessing whether the approximate construction can capture the behavior expected at a conformally invariant critical point. Its data exhibit a reasonably good collapse across different system sizes, as that in the $(1+1)$D TFIM in Fig.~\ref{fig:1d_four_point}(a). By contrast, the CDHM data show a noticeably poorer collapse and retain appreciable finite-size dependence. Although this approximate construction does not provide a definitive test of conformal invariance on the torus, the comparison suggests possible deviations from an emergent conventional CFT at the finite-parameter MDITE critical point.

\section{Results on the absence of off-diagonal long-range order in the 2D CDHM}\label{appx:nooff}
\begin{figure}[htbp]
    \centering
    \subfigure[]{\includegraphics[width=0.35\textwidth]{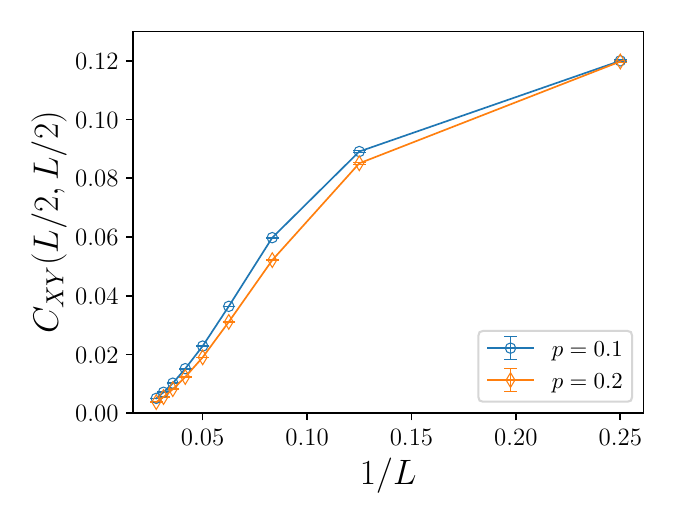}}
    \subfigure[]{\includegraphics[width=0.35\textwidth]{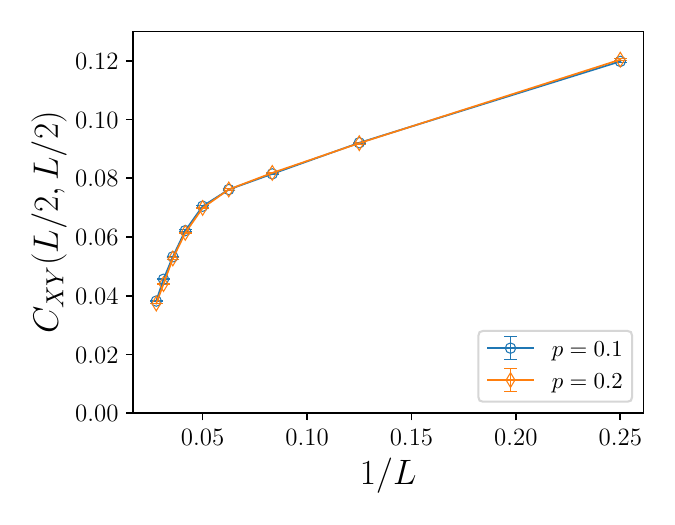}}
    \caption{
        Finite size extrapolation of off-diagonal spin-spin correlation $C_{XY} (L/2,L/2)$ at different measurement probability $p$ with (a) $\tau=16$  and (b) $\tau=100$ .
    }
    \label{fig:crxy}
\end{figure}

In this appendix, we present additional numerical results demonstrating the absence of off-diagonal long-range order in the 2D CDHM under the MDITE protocol, as discussed in Sec.~\ref{sec:result:cdhm}. 
Combining the algorithm in Sec.~\ref{appx:qmc3} and the QMC tomography method~\cite{mao2025sampling,hari2025magic,chincholi2025detecting}, we compute the correlation function
\begin{equation}
    C_{XY}(\textbf{r}) = \langle X_{\textbf{r}} X_0 + Y_{\textbf{r}} Y_0 \rangle .
\end{equation}
To extrapolate to the limit $|\textbf{r}|\to\infty$, we evaluate $C_{XY}(\textbf{r})$ at the maximum separation $\textbf{r} = (L/2,L/2)$ for each system size $L \times L$. 
We perform a finite-size extrapolation of $C_{XY}(L/2,L/2)$ as a function of $1/L$. 
As shown in Fig.~\ref{fig:crxy}, for sufficiently long imaginary time $\tau$, $C_{XY}(L/2,L/2)$ extrapolates to zero in the thermodynamic limit ($L\to\infty$), indicating the absence of off-diagonal long-range order in the mixed states generated by the MDITE protocol.

\newpage
\bibliography{ref}

\end{document}